\documentclass[manuscript]{aastex}

\newcommand{\Teff}{\mbox{$T_{\rm eff}$}~}

\newcommand{\as}{\mbox{$^{\prime\prime}$}}

\newcommand{\ebv}{$E(B-V)$}

\slugcomment{draft of \today}

\shorttitle{Extreme UV QSOs}
\shortauthors{Luciana Bianchi et al.}

\begin{document}


\title{Extreme UV QSOs} 


\author{Luciana Bianchi} 
\affil{Department of Physics and Astronomy, Johns Hopkins University,
    3400 N. Charles St., Baltimore, MD 21218, USA}
\email{bianchi@pha.jhu.edu}

\author{John B. Hutchings} 
\affil{HIA, NRC, Canada}

\author{Boryana Efremova and James E. Herald}
\affil{Center for Astrophysical Sciences, Johns Hopkins University, 
Baltimore, USA}
\author{Alessandro Bressan}
\affil{INAF- Astronomical Observatory of Padova, Italy and INAOE,
 Tonantzintla (Puebla), Mexico}

\and
\author{Cristopher Martin}
\affil{California Institute of Technology, Pasadena, USA}

\begin{abstract}
We present  a sample of spectroscopically confirmed QSOs 
with  FUV-NUV color (as measured by GALEX photometry, FUV band: 1344-1786\AA, NUV band: 1771 - 2831\AA)
bluer  than canonical QSO templates and than the majority of known QSOs.
We analyze their FUV to NIR colors, luminosities and optical spectra.
 The sample includes a group of 150 objects 
at low redshift (z $<$ 0.5),  and a group of 21 objects with redshift 1.7$<$z$<$2.6. 
For the low redshift objects, 
the ``blue'' FUV-NUV color may be caused by
enhanced  Ly$\alpha$  emission, since Ly$\alpha$  transits the GALEX FUV band from z=0.1 to z=0.47.  
 Synthetic QSO templates constructed with 
 Ly$\alpha$ up to 3 times stronger than in standard templates match the observed UV colors of our low redshift  sample.
Optical photometric and spectroscopic properties of these QSOs are not atypical.
The H$\alpha$ emission increases, and the optical spectra become bluer, with increasing absolute UV luminosity. 
The lack of selected objects at intermediate redshift  is consistent with 
the fact that for z=0.48$-$1.63, Ly$\alpha$ is included in the GALEX NUV band, 
making the observed FUV-NUV  redder than the limit of our sample selection. 
The UV-blue QSOs at redshift $\sim$2,  where the GALEX bands sample  restframe  
$\approx$450-590\AA~(FUV)  and $\approx$590-940\AA~(NUV),
are fainter than the average of UV-normal QSOs at similar redshift in NUV, while they have comparable luminosities
in other bands. Therefore
we speculate that their observed FUV-NUV color may be explained by 
a combination of steep flux rise towards short wavelengths  and dust absorption below the Lyman 
limit, such as from small grains or crystalline carbon (nanodiamonds).
The ratio of Ly$\alpha$ to CIV could be 
measured in 10 objects; it is higher (30\% on average) than for UV-normal QSOs, and close to the value expected for shock or
collisional ionization. However, optical spectra are taken at different times than the UV photometry, which
may bias the comparison if lines are variable. 
 These QSO groups are uniquely set apart by the GALEX photometry 
within larger samples,
given that their optical properties are not unusual.

\end{abstract}




\section{Introduction.}
\label{s_sample}
 In all QSO samples there is concern that selection effects are present and
significant, particularly in whether whole classes of objects are not included, or even
known. This study aims at characterizing a population of
objects with rising
fluxes at UV observed wavelengths.
Following our work of classification of  UV sources from the GALEX \footnote{
The {\it Galaxy Evolution Explorer}, GALEX (Martin et al. 2005), is a NASA Small Explorer 
performing imaging surveys of the sky in two UV bands simultaneously:
FUV (1344 - 1786 \AA, $\lambda$$_{eff}$ = 1528 {\AA})
and NUV (1771 - 2831 \AA, $\lambda$$_{eff}$ = 2271 {\AA}) with different coverage and depth. See Bianchi (2008) for a summary of the UV sources
classification and 
statistics in the main surveys and Morrissey et al. (2007) for instrument description and performance.}
 sky surveys
(Bianchi 2008, Bianchi et al. 2006, 2007, 2008 and references therein), we have suspected
the existence of a substantial number of
 extragalactic objects with FUV-NUV color  
much bluer (more negative) than  canonical QSO templates and than
the majority of  QSOs in known samples.
 Such objects are rather ``normal''  at optical wavelengths 
(spectroscopically and photometrically) but they stand out in the observed UV range,
 having FUV-NUV colors similar to those of hot white dwarfs (WD).
Photometrically, these objects have UV-to-optical colors similar 
to a stellar binary containing a hot WD and a cooler companion.   
That a significant number of  ``FUV-NUV''-blue extragalactic objects existed 
was first suspected by Bianchi et al. (2007), 
based on density counts of photometrically selected WD candidates. In fact,
the number of objects per square degree whose SED (FUV to near-IR) is consistent
with a single hot WD increases with magnitude down to m$_{UV}$$\sim$21 (AB)
and then declines, consistent with Milky Way models. However, the density of
objects with similarly blue UV color but redder optical colors, that we would expect to be  
 hot WDs with a cool companion, increases considerably  
 at fainter magnitudes, suggesting that a significant number of  faint extragalactic
objects may be included in the color-color {\it locus} of these stellar binaries
(Bianchi et al. 2007, Bianchi 2008).
In this work we focus on these QSOs, 
which display very  blue  observed FUV-NUV colors, and 
 investigate whether their properties are unlike those of known objects.

\section{Sample and data.}
The sample was extracted from the catalog of matched UV/optical sources of Bianchi (2008),
obtained by matching the UV sources in the GALEX third data release 
(GR3)\footnote{GR3 is available from the MAST archive at http://galex.stsci.edu  }, to the Sloan Digital Sky Survey (SDSS) 
sixth data release (DR6). 
GALEX provides sky surveys with different sky area coverage and depth: we restricted 
this work to the ``Medium Imaging Survey'' (MIS) data, which reaches a typical ABmag of $\approx$22.7 
in both FUV and NUV. The overlap area between GALEX-GR3 MIS data and SDSS-DR6  is 573 square degrees (Bianchi 2008),
taking into account that only the central 1~degree diameter part of the GALEX fields was
used in our master catalog, to assure homogeneous photometry quality and exclude defects 
in the outer parts of the circular field. 
For each matched source, GALEX provides FUV (1344 - 1786 \AA, $\lambda$$_{eff}$ = 1528 {\AA})
and NUV (1771 - 2831 \AA, $\lambda$$_{eff}$ = 2271 {\AA}) photometry, and the SDSS provides {\it u, g, r, i, z} photometry.  
More details on the matchings procedure and the catalog are given by Bianchi (2008), Bianchi et al (2009). 

In order to characterize the suspected ``FUV-NUV blue'' QSOs, we 
extracted from the matched UV/optical source catalog of Bianchi (2008) the
spectroscopically confirmed QSOs with  
 FUV-NUV $<$ 0.1 (AB mag): this FUV-NUV limit
is ``bluer''(more negative) than the synthetic FUV-NUV color from the two QSO canonical 
templates used by Bianchi et al. (2007), which represent average QSO properties,
at any redshift.  The colors of the canonical templates are 
shown in Fig. \ref{f_ccd} and \ref{f_ccd2} (cyan diamonds) as a  function of redshift.
 We will refer to this sample as ``UV-blue'' QSOs for brevity throghout the paper. 
 It is restricted to sources with photometric errors smaller than
0.3~mag in both FUV and NUV, and  color   FUV-NUV $<$ 0.1, 
for which SDSS spectroscopy exists and gives a ``QSO'' classification.
 The requirement of available SDSS spectroscopy effectively  limits 
the sample to brighter magnitudes,  but it provides a classification and useful information,
 which will help interpreting larger samples of photometric candidates.
Such relatively bright objects will also be accessible to the spectroscopic capabilities of the
refurbished HST, and to other follow-up observations. 
Note that the SDSS spectroscopic class ``QSO'' (class 4) probably includes also Seyfert galaxies.
 We will use here the generic  term ``QSO'' to
reflect our selection criterion  from the SDSS spectroscopic database. 
It is important to note that spectroscopic targets in the SDSS were selected with criteria 
unrelated to our present objective and therefore our UV-blue spectroscopically confirmed QSOs may be a 
biased sub-sample among the UV-blue QSO photometric candidates. 
 
 These selection criteria produced an initial sample of 174 objects. 
One additional object was excluded because its u-band measurement is saturated. 
The photometric properties of the sample QSOs are presented in 
section \ref{s_photo}, and their optical spectra and overall properties
are analyzed in section \ref{s_spectra}.
The selected objects are shown in two color-color diagrams, Figs.
\ref{f_ccd} and \ref{f_ccd2}, where they can be compared with other classes of objects, 
in particular hot stars, typical QSOs, and galaxies.  Our analysis of the spectra (section \ref{s_spectra})
generally confirms the redshift measurement from the pipeline. However, we found that one object 
(GALEX J172101.08+532433.7,  RA=260.2544916,  Dec=53.4093516, SDSS match id= 587725490527731868)
was misclassified by the SDSS
spectroscopic pipeline as a QSO with redshift z=2.7: its spectrum is that of a hot star. It is 
shown in some figures because it is interesting to note its position in the color-color diagrams (Figs \ref{f_ccd} and \ref{f_ccd2}):  the
GALEX photometry clearly place this object on the stellar sequence and not as a QSO candidate.
Coordinates, photometry, and other relevant information are given in Table \ref{t_sample}, and sample images
are shown in Fig. \ref{f_fchart_new}.

The SDSS optical spectra (range $\sim$ 3800-9200\AA, resolution $\sim$1800), provide the initial classification as QSOs and 
a measure of redshift for the objects.
Our sample includes a group of 151 objects at low redshift (0.041$<$z$<$ 0.436),  
 and  21 objects with redshift between 1.7 and 2.6, all pointlike.
 Only one object has intermediate redshift (z=0.93) and its identification as a QSO is dubious.
It has fairly large photometric errors:
FUV=21.49$\pm$0.12, NUV=21.43$\pm$0.10, but the typical QSO FUV-NUV color at this redshift
is much redder (by $>$1 mag, see Fig. \ref{f_ccd}); therefore if it is a QSO it would be quite anomalous. 
At this redshift the GALEX bands sample rest wavelengths of $\sim$1300\AA(NUV)  and $\sim$800\AA(FUV), and
a very blue FUV-NUV color would be not expected. 
The image and spectrum of this object are  shown in Fig. \ref{f_oddball} -bottom. 
The spectrum  shows one emission line that is identified as MgII for the alleged redshift and
possibly a few absorptions including one at the red end that could be H$\gamma$. 
The observed wavelengths of the lines are
not obvious for identification, assuming other values of  redshift the observed line(s) might be
CIII] $\lambda$1909 or [OII]$\lambda$3727,  but then other lines such as CIV~$\lambda$1550 or [OIII]~$\lambda$5007 should be 
present and are not.
So, either z=0.93 is right or perhaps the emission is some artifact in the spectrum.
Therefore, we consider this object doubtful and do not include it in our analysis.
The lack of objects between redshift 0.5 and 1.7 is consistent with our selection of
very blue FUV-NUV color, because Ly$\alpha$ is in the NUV band between z=0.48 and z=1.63, 
causing brighter flux  in NUV and consequently much redder FUV-NUV color,
as can be seen in Fig. \ref{f_ccd}. 

We also caution that while the GALEX FUV and NUV images are taken simultaneously,
the SDSS imaging was taken at a different time from the
GALEX observations, therefore  any significant variability may affect the 
combined UV and optical colors, such as NUV-r. For this reason we based our initial sample
selection on the FUV-NUV color only. Some of our targets have repeated observations with GALEX,
but most repeated measurements are from the AIS (All-sky Imaging Survey), which has about 10 times shorter exposures
than MIS (used in this work) and therefore large photometric uncertainties.
In a few cases repeated measurements are discrepant by $>$2$\sigma$ in the combined photometric errors: 
however most of the discrepant measurements have artifact flags set. 
We compile for completeness
 all repeated measurements with exposures longer than 400~sec and  formally discrepant by  $>$2$\sigma$
in Table \ref{t_dup}, where we provide also comments that help assess reliability, based on the flags 
from the pipeline photometry, and our visual inspection of the images. 
We only excluded measurements on the very edge of the GALEX field (flag ``rim'') however we 
did not apply error cuts nor area cut, for the purpose of an exhaustive comparison,  while our
analysis sample is restricted to measurements in the central 0.5~deg. radius of
the field for accurate photometry (Sect. 2).   
In a few cases the discrepancies in the repeated measurements
 cannot be ascribed to artifacts, and these objects may deserve dedicated follow-up photometry.  
In some cases the variation affects the FUV-NUV color, and in particular some repeated measurements 
have redder  FUV-NUV than our initially selected dataset. 
All discrepant repeated measurements with MIS exposures (2 high-redshift objects 
and 10 low-redshift objects) have FUV-NUV$>$0.1.  If we consider also AIS data 
(exposure times $\sim$ 100~sec), we find 55 additional 
repeated measurements discrepant by $>$2~$\sigma$,
of which 36 give FUV-NUV redder than our selected dataset (MIS measurements), and other 19 bluer.
 Fast variability is not unknown in QSOs,
and in particular line strength may vary on short time scales, while it would
be less plausible for dust effects to change rapidly. We stress that a variabilty assessment however
would require custom photometry, and while the standard pipeline photometry is good for statistical analysis,
such as the scope of this work, 
we should refrain from overinterpreting individual measurements and individual variations. 
Other 72 AIS and 7 MIS repeated measurements agree within 2$\sigma$ with our selected measurements given in Table 1. 
One object in the initial sample, although part of the MIS survey, has a 40~sec exposure 
in FUV and 1518~sec in NUV: although its
FUV error (0.23) meets our selection limits, a longer MIS exposure of 853~sec in both
 FUV and NUV and smaller errors gives FUV-NUV=0.17, so it 
is eliminated from our analysis sample. 

A larger sample of about 30,000 QSOs candidates  with ``normal'' FUV-NUV colors
(i.e. similar to the standard template), 
will be presented elsewhere. We will refer to this sample as ``UV-normal'' 
 in the discussion of the UV-blue sample  for comparison purposes.

\section{General photometric properties}
\label{s_photo}

Of the 174 sample objects, 64 sources are classified as point-like (at the resolution of the SDSS 
imaging, $\sim$1.4$^{\prime\prime}$) and 110 as extended (all at low redshift), by the SDSS pipeline.
 We will keep the pipeline classification because it is derived from an objective procedure, although the result depends on
the contrast between central source and underlying galaxy. GALEX and SDSS imaging for 
a subsample of objects, presented in Fig. \ref{f_fchart_new},
shows that the definition of ``pointlike'' (P) or ``extended'' (E) is not clear-cut.
 
 The sample selection,  
as described in  section \ref{s_sample},
was restricted to MIS sources with  photometric errors less
than 0.3~mag in FUV and NUV.    Of the 64 point-like
sources, 42/33 have  errors less than 0.1~mag/0.05~mag,  and only 4
have errors between 0.2 and 0.3~mag, in FUV. As for the NUV measurements,  
36/45 pointlike sources have errors $<$~0.05/0.1~mag, and 10 have errors
larger than 0.2~mag. In the {\it r-}band, all but 2 objects have errors smaller than
0.05~mag  (one object has an error of 0.14~mag, and one of 0.08~mag).
Of the 110 extended objects, 108 have {\it r-}band error $<$0.04~mag,
97/94 have NUV / FUV error $<$ 0.1~mag. Sources with large photometric errors  
are identified in the figures. 
Most objects with the larger errors are in the z$\sim$2 group. At this redshift, our color
cut of FUV-NUV$<$0.1 is more than half a magnitude bluer than the average value for UV-normal QSOs (e.g. Fig.1), 
therefore these objects may be truly extreme with respect to average samples in spite of their large photometric errors. 
We point out that the object GALEX J113223.4+641958 (SDSS J113223.42+641958.4) has a u-band magnitude of 
28.7$\pm$0.46 (petromag) while the magnitude listed on the explorer page for this object is u=25.07$\pm$3.05.
It has no artifact flags, the pipeline records only a warning ``no petrosian radius could be determined. Petrosian magnitude
still usable; the object is blended with an extended object''. The surrounding galaxy can be seen in the SDSS
imaging with a radius of about 5$\as$. We regard the petrosian magnitude as unreliable in the u-band.
Magnitudes in other bands for  this object have smaller errors and seem more consistent among measurements.   
All other objects have u-band magnitudes brighter than 22.2, consistent with the SDSS limit (see Fig.~3 of Bianchi et al. 2007). 
We give in Table \ref{t_sample} petrosian magnitude measurements for the SDSS data, for consistency 
among the sample and with other extragalactic works.  
The SDSS pipeline also provides magnitudes measured in different ways: psf fitting, 
DeVaucouleurs model, and exponential fitting. A description of the different magnitudes 
can be found on the SDSSS web site  http://www.sdss.org/dr5/algorithms/photometry.html. 
 We checked for all objects whether the different 
measurements are discrepant. As expected, for pointlike sources the average difference is 
within the 1$\sigma$ errors and the largest discrepancies close to 3$\sigma$.
 Disagreement between psf-mag and petromag tend to increase at longer wavelengths, where the
extended galaxy is contributing. 
For extended objects, the measurements from petrosian and deVaucouleur-profile fitting agree on average within
better than  2$\sigma$, while psf magnitudes are more discrepant as expected and should not be used.  

  The FUV-NUV and NUV-g colors of the sample objects are plotted as a function of redshift in Fig. \ref{f_uvcolor},
and the FUV, NUV, and {\it r-}band magnitudes in Fig. \ref{f_mags}. 
 ``Extended'' sources are plotted with different symbols, to explore possible trends, although the classification
must be regarded only as an indication as pointed out above.  
Photometric errors (1~$\sigma$ error bars are shown) 
in most cases are quite small compared with the spread in FUV-NUV  color
observed in our sample. 
Fig. \ref{f_uvcolor} shows that the high-redshift objects have
a wider range of FUV-NUV color than the low-redshift point-like sample, although the spread 
may simply be caused by the large errors of these faint objects. 
The lack of extended objects at high redshift is likely due to the fact that for these
more distant objects the same imaging does not reveal the underlying galaxy.  
This question will be investigated with deeper
imaging aimed at revealing the underlying galaxy in the distant objects and to probe the contrast to the 
central source (Hutchings, Scholz, and Bianchi 2009). 

Fig.  \ref{f_mags} shows that low-redshift pointlike  QSOs tend to be  brighter than 
extended ones,  in both FUV and NUV.
 In the {\it r-}band, however,
the magnitude spread is less (about 4 mags across the sample) and no preferential distribution is 
seen between pointlike and extended samples. Low redshift pointlike objects 
are also brighter (observed magnitudes) than 
higher redshift objects by about 2-3 magnitudes, but their intrinsic luminosity is lower.
 The distribution of observed magnitudes (left panels)  is useful for comparison with other samples,
and to estimate the possible contamination by these objects to density counts of 
other UV-blue objects such as Milky Way  WDs, which have similar FUV-NUV colors (see Bianchi et al. 2007, 2008,
and Fig. 1),  as well as for planning follow-up observations. 
The misclassified star is shown in these panels. In the right-side panels of Fig.  \ref{f_mags}
the absolute magnitudes are plotted (the luminosity distance was derived from the redshift
using standard cosmology  H$_0$ = 70 km/s/Mpc, $\Omega_M$=0.3, $\Lambda$ = 0.7 );  
 the distant objects are intrinsically more  luminous. 
 We plotted for  comparison the median absolute magnitudes
of our UV-normal QSO comparison sample (solid line in the right side plots). 
The high redshift UV-blue QSOs have luminosities similar to UV-normal QSOs, 
except in NUV, where they are fainter.
The comparison suggests some absorption in the NUV band (restframe $<$900\AA~ for z=2) as an
explanation of their FUV-NUV color. We will discuss this point later.

\section{Analysis and Discussion.}
\label{s_spectra}
We discuss the two groups, z$<$0.5, and z$\sim$2 QSOs, separately 
because the FUV and NUV bands sample different restframe spectral regions and therefore
the explanations for their blue FUV-NUV colors are different.  

\subsection{The low redshift QSOs}

The majority of our analysis sample has redshift  $<$0.5 (150 objects, 109 extended and 41 pointlike).
 Ly$\alpha$ transits the GALEX FUV band from z=0.1 to z=0.47, and
this fact suggests that an intense 
Ly$\alpha$  emission may be the cause for the ``FUV excess'' of 
these objects.   To test this hypothesis, we  constructed templates with 
 Ly$\alpha$ emission enhanced relative to standard templates, 
and derived their synthetic broad-band colors. 
 Such {\it ad hoc} templates
with Ly$\alpha$ enhanced by up to  3$\times$  match the range of observed FUV-NUV colors of our low-z sample, and
are shown in Figs.  \ref{f_ccd} and \ref{f_ccd2} (dark blue diamonds) together with
synthetic colors from canonical templates (cyan diamonds), as well as in Fig. \ref{f_uvcolor}(top).
Note from these figures that our simple cut of FUV-NUV$<$0.1 produced a low-redshift sample
bluer in FUV-NUV than standard templates with a spread of about half a magnitude: the redder objects 
among our sample are very close to the standard template
at z$\sim$0.2, the UV-bluest objects differ by up to .5~mag (one by $\sim$1~mag) and
are concentrated around z$\sim$0.2  where Ly$\alpha$ is at the peak of the FUV filter
transmission. 
The modulation with redshift of the hypothetical enhanced-Ly$\alpha$ effect, 
due to the filter transmission,  is seen in the {\it ad hoc} template plotted in Fig. \ref{f_uvcolor}.

None of our sample objects have UV spectra, which would directly reveal the cause of their
blue FUV-NUV color. 
We examined their optical spectra and in particular H$\alpha$, the strongest line in all the objects. 
   Fig. \ref{f_sp_stack_loz} shows the optical spectra
of our low redshift sample, stacked, and compared with the standard template (cyan). 
The majority of the pointlike sources have  
emission lines stronger than the average template, and bluer spectral
slope (flux increasing at shorter wavelengths). For the  extended sources, however, line strength is 
generally typical and  the spectral slope mostly redder than the standard template,
 reflecting the non-negligible contribution by the underlying galaxy (SDSS
spectra are taken through a 3\as ~ diameter aperture). Sample spectra 
 for both pointlike and extended QSOs are also shown in Fig. \ref{f_sp_loz}.
 There is a wide range of line
strengths and profiles, as well as spectral slopes. 

The SDSS pipeline provides automated measurements of width and equivalent width (EW) of the major lines,
performed with line fitting; 
we downloaded and examined those quantities. We found that the centering of the line could be
used, while line width and equivalent width from the pipeline are not reliable for most
spectra  (examples in Fig. \ref{f_hawidth}). We remeasured the  H$\alpha$ line, 
first by hand to assess the difference from the pipeline measurements,
and then with an {\it ad hoc} algorithm for  more objective results.
The line width estimated by our code is also shown in  Fig. \ref{f_hawidth}. In order to minimize
 the complication of narrow absorptions and emissions in some profiles, 
we did not measure the width at half maximum (peak) but the width at the average flux value of the 
total line emission.
 We consider our measurements more homogeneous than the pipeline values, 
as shown in Fig. \ref{f_hawidth}, and we use them in the following analysis.
 H$\alpha$ width, EW and fluxes (F$_{\lambda}$) measured (at restframe wavelengths) 
for the low-z sample are reported in Table \ref{t_meas}. 
Errors from the spectra S/N and continuum placement uncertainties, are estimated to be less than 10\%. 
As a further check, measurements from our code agree with our manual measurements with 
by-eye location of the continuum to a few percent in all but a few cases, where they agree within 10\%. 
Our measurements include the entirety of the emission feature, no attempt was made to
separate narrow components when present, and no correction for [NII] was applied. 

 We searched for possible correlations of H$\alpha$ intensity and width, and of the optical spectral slope, 
with the UV color and absolute luminosity. 
The spectral slope was measured as the ratio of fluxes integrated  in two intervals
which are rather free of conspicuous features in most spectra: 
rest wavelengths 3500-3700\AA~ and 6000-6400\AA. 
We compared several quantities, and we show six interesting cases in Fig. \ref{f_ha}. 
No obvious correlation is seen with the FUV-NUV color. 
The  H$\alpha$ flux,  and EW, increase with absolute UV luminosity, 
and to a much lesser extent with u-band luminosity,
but not with luminosity at longer wavelengths. 
Similarly, the optical spectral slope %
possibly correlates with H$\alpha$ EW and with UV luminosity
(but not with optical luminosity: the r-band is also shown in Fig.\ref{f_ha}):
it becomes bluer for brighter UV luminosities, suggesting 
 that for low QSO luminosity the galaxy relative contribution is more significant. 
 There is a clear difference between pointlike and extended sources:
the latter have a flatter (redder) optical slope and lower H$\alpha$ emission, 
reflecting the contribution of the host galaxy.  
 This result emphasizes the role of UV studies in extending the known properties of QSOs.
If we restrict the sample around redshift z=0.2, 
where we have a wider FUV-NUV observed range and Ly$\alpha$ is at the peak of the filter's transmission,
the scatter is much reduced in the correlations with absolute FUV luminosity, and 
 some possible correlations with UV color  emerge, but the number of points is then
too scarce for robust conclusions.  
Alternative explanations for the blue FUV-NUV color
 may include a dust effect, depressing the NUV flux. However, 
it is not obvious that any known interstellar
extinction law would have this effect at these redshifts: the 2175\AA~ dip,
for instance, would lie at the upper (long wavelength) end of the NUV passband for redshift beyond 0.2, and the FUV would
be more absorbed.  There is no correlation of the H$\alpha$ intensity with foreground E(B-V).

Figure \ref{f_sedLowZ} shows the magnitudes of the low-redshift sample, and their median values
connected by  lines. The average SED of  UV-normal QSOs in this
redshift range is also shown for comparison. The extended sources differ from the pointlike ones. 
Among pointlike sources, the overall brightness of UV-normal QSOs is slightly 
lower than the UV-blue QSOs.
The extended UV-blue  and UV-normal  samples have similar SED in the optical bands, showing similarity of the host galaxy
which contributes to the flux. 
We performed this comparison both using dereddened magnitudes, where each photometric measurement was 
dereddened using E(B-V) (Table 1) estimated from  the Schlegel et al. (1998)  maps,
before averaging the sample, as well as using observed magnitudes without extinction corrections.
The individual sources and the average SEDs shift correspondingly, by up to $\sim$0.4mag in UV,
but the relative differences among average SEDs remain the same. 

While there is no UV spectroscopy for our objects, we examined UV    HST-STIS archival spectra of a
 sample of QSOs %
published by Shang et al. (2005). None of them are included 
in the current GALEX MIS coverage, but a few are in the GALEX AIS 
(which has about 10 times shorter exposures than MIS data
and therefore larger photometric errors).
We also computed synthetic colors
for the Shang et al sample convolving the observed HST/STIS spectral fluxes with the GALEX transmission bands.
We show their position on the color-color diagram (Fig. \ref{f_ccd}, small teal diamonds).
A few of the Shang et al (2005) QSOs have 
FUV-NUV only slightly bluer than 0.1, %
while the rest have  typical FUV-NUV colors. 
The QSOs in the STIS sample with FUV-NUV $<$0.1 have optical spectral slope and H$\alpha$ emission similar to the standard 
QSO template, but most display stronger Ly$\alpha$ and CIV, and a range of UV slopes
(steeper, similar, but also flatter, than the average).
The comparison, although limited to a different sample than ours and to a few bright objects, suggests
that unusual UV properties may exist that cannot be predicted from the optical data.

\subsection{The high redshift QSOs}
\label{s_hiz}

The photometric SEDs of the 21 UV-blue QSOs with redshift  between 1.7 and 2.6 
are shown in Fig. \ref{f_jh1},  and their optical spectra in Figs. \ref{f_sp_hiz} and \ref{f_sp_stack_hiz}. 
The QSO with the highest redshift in the sample (z= 2.6) 
has an  extremely red optical spectrum and appears as a very red faint object
in the optical imaging (Fig. \ref{f_oddball}).
It  also has much larger photometric errors than the rest of the sample:
FUV=22.41$\pm$0.19 , NUV=22.67$\pm$0.29,  
{\it u}=22.22$\pm$0.8, {\it g}=21.40$\pm$0.23, {\it r}=20.86$\pm$0.14.
The typical QSO FUV-NUV color for 
its redshift  is significantly redder (more than half magnitude, Fig. 1)
so the object may still deserve attention. 

At redshift z=2, the NUV band includes flux in the restframe range
590 - 940\AA~ (the  filter's $\lambda$$_{eff}$ becomes restframe  757\AA) 
 and the FUV band includes restframe 450-590\AA ~($\lambda$$_{eff}$$\sim$ restframe  500\AA).
We speculate that a combination of steep flux rise towards restframe extreme ultraviolet (EUV) and absorption below the Lyman 
limit may explain the observed FUV-NUV color.
We constructed spectral templates with FUV flux rising more steeply than in standard templates, using
 two power-law slopes F$_{\lambda}$ $\sim$ $\lambda$$^{\alpha}$ with $\alpha$=-0.6 and -1.2.
The average slope between 500$--$1200\AA~ in the large sample of Telfer et al. (2002) is
F$_{\nu}$ $\sim$ $\nu$$^{-1.76}$, i.e. ${\alpha}$=-0.24 in  F$_{\lambda}$. 
Our EUV-steep templates are shown with dark green diamonds 
in Figs. \ref{f_ccd} and \ref{f_ccd2}. While they  have synthetic FUV-NUV color bluer than the  canonical template
at redshift $\sim$ 2,  they are still more than half magnitude redder than the colors
observed in our UV-blue sample.  The fact suggests that a combination of both EUV flux rise 
at shorter wavelengths and a deep absorption
below the Lyman limit may be required to explain the observed colors of our UV-blue QSOs. 
The suggestion is supported  
by Fig. \ref{f_mags}, showing the absolute NUV luminosity of our UV-blue sample to be lower
than average.  This can also be appreciated in 
Fig. \ref{f_jh1} which  shows all the observed magnitudes for our high-redshift sample. The Lyman
limit in this redshift range lies between the NUV and u bands, and the
Lyman drop is clearly seen. The line shows the median values, and 
 only the object with a red optical spectrum mentioned above differs 
significantly (shown by the dotted line). 
 Figure \ref{f_jh1} also shows the median magnitudes for UV-normal QSOs 
from the MIS survey, with the same redshift range and error cuts. 
The average FUV-NUV is much `redder' for the UV-normal QSOs, consistent with our selection. 
The UV-blue QSOs are fainter in the NUV band, which is sampling
the Lyman limit at these redshifts. Thus, our QSOs sample may have somewhat more
extinction and more severe absorption below the Lyman limit.
 Three of the UV-blue QSOs have strong BAL-type C IV absorption (Fig. \ref{f_sp_hiz}). 
Their Lyman discontinuities (estimated from
the broad-band photometry, as defined in Fig.\ref{f_jh2})
are very large for two of them and smaller than average for one. Thus, it
is not clear whether BAL absorbers contribute signifcantly to the Lyman drop. 

 Binette \& Krongold (2008, and references therein) discuss the spectrum of Ton~34, 
an unusual QSO with %
an enhanced ``Lyman valley'' in its UV spectra (IUE and HST), which can be reproduced by their models of 
absorption from carbon crystalline dust (nanodiamonds). Ton~34 is at  redshift z=1.93 
and  we investigated  whether it could be a possible counterpart of our UV-blue QSOs.  
It is not included in the GALEX 
 surveys to date (it is just outside the edge of an observed  GALEX field), so
we estimated GALEX FUV and NUV magnitudes by convolving the  IUE SW and LW spectra of Ton~34
with the GALEX filters, and  obtained FUV-NUV $\sim$1.3,
close to the expected color of the UV-normal sample at this redshift and much
redder than our UV-blue QSOs. This color estimate is uncertain because in the IUE spectra the 
signal is close to the background limit, and 
HST spectra of Ton~34 do not even cover one of the
GALEX bands. %
 The GALEX FUV band includes flux longwards of 1344\AA, 
while in the IUE spectrum of Ton~34 the flux is very steeply rising just shortwards of this limit. 
Therefore, a slightly more
redshifted analog of Ton~34 would produce a much brighter FUV magnitude. 
 
Models of crystalline dust absorption by Binette \& Krongold (2008) show that the 
general effect is a very deep Lyman valley, and in more detail the relative amounts of absorption in the
wavelength ranges sampled by the GALEX FUV and NUV filters at z$\approx$2 vary according to
the dust geometry and composition. 
For example, comparison of dust models in figure A.2 of Binette \& Krongold (2008) suggests that
a lower column density of the carbon crystalline dust screen (or an intrinsic SED steeper towards short wavelengths)
may produce a higher FUV flux, and  small grain dust (similar composition as Milky Way
dust, i.e. silicate and graphite grains, but grain sizes much smaller than MW dust and larger than
nanodiamonds) would cause a significant depression of the observed-NUV flux but less reduction of the
observed-FUV at the redshift  of our high-z UV-blue QSOs. 
This effect would be qualitatively consistent with the SED of our UV-blue QSOs. 
From broad-band photometry alone, it is not possible to separate effects of
dust absorption and intrinsic SED slope, therefore %
we can only speculate that the observed FUV-NUV colors in our sample are 
qualitatively compatible with absorption from dust with grains
differing from Milky Way dust (smaller grains), and possibly a steeper flux rise towards EUV.  
 The question remains open as to what causes the extremely blue FUV-NUV colors, and whether
these objects have known counterparts with similar properties, until UV spectroscopy
can be obtained.  

We have measured the emission lines of C IV and C III] (EW, total flux, and full width (FW)
at 10\% of the peak flux above the local continuum)
 from the SDSS spectra. Typical errors are of the order of 10\%.
Ly$\alpha$ is generally too near the 
end of the optical spectra and could be measured only in ten cases. The C III] line is free of absorptions
but some QSOs have significant BAL and interstellar absorptions in the CIV line. 
The emission line properties do not correlate at all with the FUV-NUV color. 
The FUV-NUV color does correlate with the g-i color, which may indicate
that extinction is involved, and  with
redshift, although weakly, in the sense that higher redshift objects are more
UV-blue (Fig. \ref{f_uvcolor} and \ref{f_jh2}). 
This is what we would expect in the rest wavelengths below the
Lyman limit, where the continuum is rising again. The Lyman discontinuity 
is larger for higher redshifts too, which is likely caused by where it lies
between the NUV and u bandpasses. 
   The emission line EW is larger for fainter FUV magnitudes, but scales more
slowly than the continuum flux. The line full-width is higher for more luminous
QSOs, based on their g-band magnitudes (rest frame FUV). 
Flux and EW of C III] and C IV lines correlate with the Lyman discontinuity, but not the line full 
width. Thus, there is a connection between line emission and the EUV continuum. 
Figure \ref{f_jh2} shows some of these correlations; the Spearman's ${\rho}$
significance test gives a probability of correlation (clockwise from top left)
of 99\%, 94\%, 58\% and 99\%. 

While the  dust absorption  affects more the continuum, 
the ionization would be reflected by  the line ratios.
 Binette \& Krongold (2008, and references therein) discuss also 
the effects of shock ionization versus photoionization. 
 It is interesting that their models show low C~IV and N~V relative to Ly$\alpha$, compared with UV-normal QSOs.
 We  measured  Ly$\alpha$$+$NV and C~IV in our high-redshift UV QSOs where possible (10 cases).
The line flux ratios may be useful diagnostic since shocks may not be
related to the continuum. %
 Therefore, we also examined SDSS spectra of UV-normal QSOs in the same redshift range 
and compared their line strength with the UV-blue sample. We extracted spectroscopically confirmed
QSOs in the same redshift range z= 1.7-2.5, but with FUV-NUV $>$0.1, from our master catalog of matched sources. 
We found 420 objects, compared with our 21 with FUV-NUV $<$0.1. 
We imposed the same error cuts in FUV, which unfavours
the red (normal) QSOs, so the ratio (5\%) is a lower limit for the
fraction of UV-blue QSOs compared with normal ones.
The relative numbers however  may be highly biased because the SDSS spectral targets
were chosen with criteria not related to our UV selection. %
We measured the same line ratio only for the UV-normal comparison objects with z=2.2-2.5, where Ly$\alpha$ is included
in the optical spectra.  The measurements are shown in Fig. \ref{f_ratio}, where a linear
fit is also shown; the formal probability of correlation is over 99\%.
 The average  line ratios for the UV-blue and UV-normal samples are  given in Table \ref{t_ratios}. 
The average is 5.2 for our UV-blue sample and 3.7 for our normal comparison sample.
The collisional model predicts a ratio of $\sim$6.7, and the photoionization model 1.8.
In one of our UV QSOs the ratio Ly$\alpha$+NV~/~CIV is about 10 while Ton~34 has a ratio of  8.7.
This bears out the similarity with Ton~34, and a dominance of collisional ionization, compared with 
UV-normal QSOs.

\section{Conclusions and summary}
  We analyzed 171 spectroscopically confirmed QSOs with  FUV-NUV color bluer than 0.1, extracted
from the GALEX MIS survey with complemetary SDSS
optical data. Most of these objects have redshift $<$0.5, and we speculate that 
Ly$\alpha$ emission enhanced up to a factor of 3 with respect to average templates,
may explain the observed colors. %
Their optical properties are similar to those of UV-normal QSOs.  Both photometric and emission line properties differ
between point-like and extended sources, reflecting the contribution from the host galaxy in the latter.
The slope of their optical spectra and 
the strength of H$\alpha$ (flux and EW) correlate (increase) with intrinsic UV luminosity.
 Ly$\alpha$ goes through the GALEX FUV band in the
redshift range of these objects, between 0.1 and 0.5, therefore the resulting effect on 
the broad-band FUV magnitude is a combination of the line intensity and the filter's transmission curve. 
A restricted sub-sample with redshift around 0.2 (where  Ly$\alpha$ is at the peak of the
filter's transmission) seems to show tighter correlations but it is statistically insufficient to support conclusions.
The UV luminosity is brighter ($\sim$0.5 to 1~mag on average) than that of our
UV-normal comparison sample, the difference being larger in FUV and for the pointlike objects (Fig. \ref{f_mags} and \ref{f_sedLowZ}).

 Our sample of UV-blue QSOs  also includes  21 objects with redshift between 1.7 and 2.6. 
Their photometric errors are generally large, 
the combined FUV-NUV 1~$\sigma$ errors are between 0.1 and 0.37~mags,  but our FUV-NUV selection limit
(FUV-NUV $<$0.1)
is bluer than typical QSO colors at this redshift by more than half magnitude,
and the observed FUV-NUV colors are bluer than the typical color by up to 1 magnitude or more (Fig. 1).
For these UV-blue QSOs at higher redshift we speculate that a combination of unusually strong
absorption in GALEX-NUV (restframe $\sim$600-900\AA) and EUV-steeply rising  flux
(GALEX FUV $\sim$ restframe 450-590\AA) may explain the FUV-NUV color. 
This is suggested by two facts. First, ad-hoc templates with flux rising towards restframe EUV 
more steeply than in  
 canonical templates, produce observed  FUV-NUV colors bluer than the average template 
(Fig. \ref{f_ccd2}), but still redder than our selection limit by 0.2~mags and redder than
most of our UV-blue sample by up to 1 mag.  Second,  comparison with average SED of 
a UV-normal QSO sample, shows the 
 NUV luminosity of the UV-blue sample to be fainter, suggesting absorption in the observed NUV.
Dust with composition similar to the typical Milky Way dust but smaller grains 
and carbon crystalline nano-size grains (nanodiamonds)
would cause absorption in the observed NUV band, according to the models of 
 Binette and Krongold (2008), which may qualitatively account for the observed 
FUV-NUV colors.  UV spectroscopy %
is needed to pinpoint the cause for
the FUV-NUV color of these objects. 
 The Ly$\alpha$ to CIV  ratio is stronger in the optical
spectra of the UV-blue QSOs than in the UV-normal comparison sample (at the $>$95\% confidence
level from K-S test, although both samples are very small, see Fig. \ref{f_ratio}), suggesting
collisional ionization to be more relevant in the UV-blue QSOs.

 The group of UV QSOs at z$\sim$2 may  probe a particularly relevant phase
of galaxy formation, tighly connected with the 
formation of the massive central black hole. In current QSO/Spheroid coevolution models (e.g. Granato et al 2004)
the power of the central QSO rises almost exponentially
and quickly stops the star formation process.
During the previous phase it is strongly dust enshrouded
and not visible except in X rays.
A phase of decreasing 
(but still significant) extinction follows, and finally a shining phase until the fuel is consumed.
A quick transition is expected between dust-extinguished and
not extinguished phases for QSOs at high z.
The relative percentage of UV-blue QSOs could be a measure of the relative lifetimes
of that phase. As explained in section \ref{s_hiz}, within our sample of matched GALEX-SDSS sources,
the number of spectroscopically confirmed QSOs with FUV-NUV$<$0.1 is about 5\% of those 
with redder FUV-NUV in the redshift range around 2. However, this number may be highly biased
because the availability of SDSS spectra is serendipitous from the point of view of our selection.
The fraction of sources with available spectra is not uniform across the range 
of  optical and UV colors, and redshift, of our photometric candidate sample.  
Spectroscopic selection especially favours the brightest samples, while 
Bianchi et al. (2007, 2008) show for example a steep increase of UV-blue extragalactic object candidates at
faint magnitudes. Some UV-to-optical color ranges are also contaminated by stellar objects and 
the purity of photometric candidate samples varies greatly according to the colors regime and parameters. 
The aim of this work was to point out that a non-negligible sample of UV-blue QSOs exist, 
and explore their nature. Statistical considerations will be addressed using a larger sample. 

 Another possible bias may arise from  variability, which is frequently
observed in QSOs. Serendipitous repeated UV observations for our sample show 
variations by $>$3$\sigma$ in some objects, and in many cases the FUV-NUV in repeated measurements
is redder than in our selected data-set, making some of these objects UV-normal or close to
normal in some of the measurements, and extremely blue in others. We have tried to exclude as
thoroughly as possible imaging or pipeline artifacts, using the flags provided by the pipeline and
identifying several additional unreliable measurements by individual analysis. However, we should
keep in mind that pipeline photometry of large datasets has statistical value, and in particular
the combination of GALEX and SDSS source catalogs over a large area of the sky proved invaluable to
characterize elusive classes of objects (Bianchi et al. 2007),   but it should not be overinterpreted
for individual objects.

\begin{figure} 
\includegraphics[scale=.65]{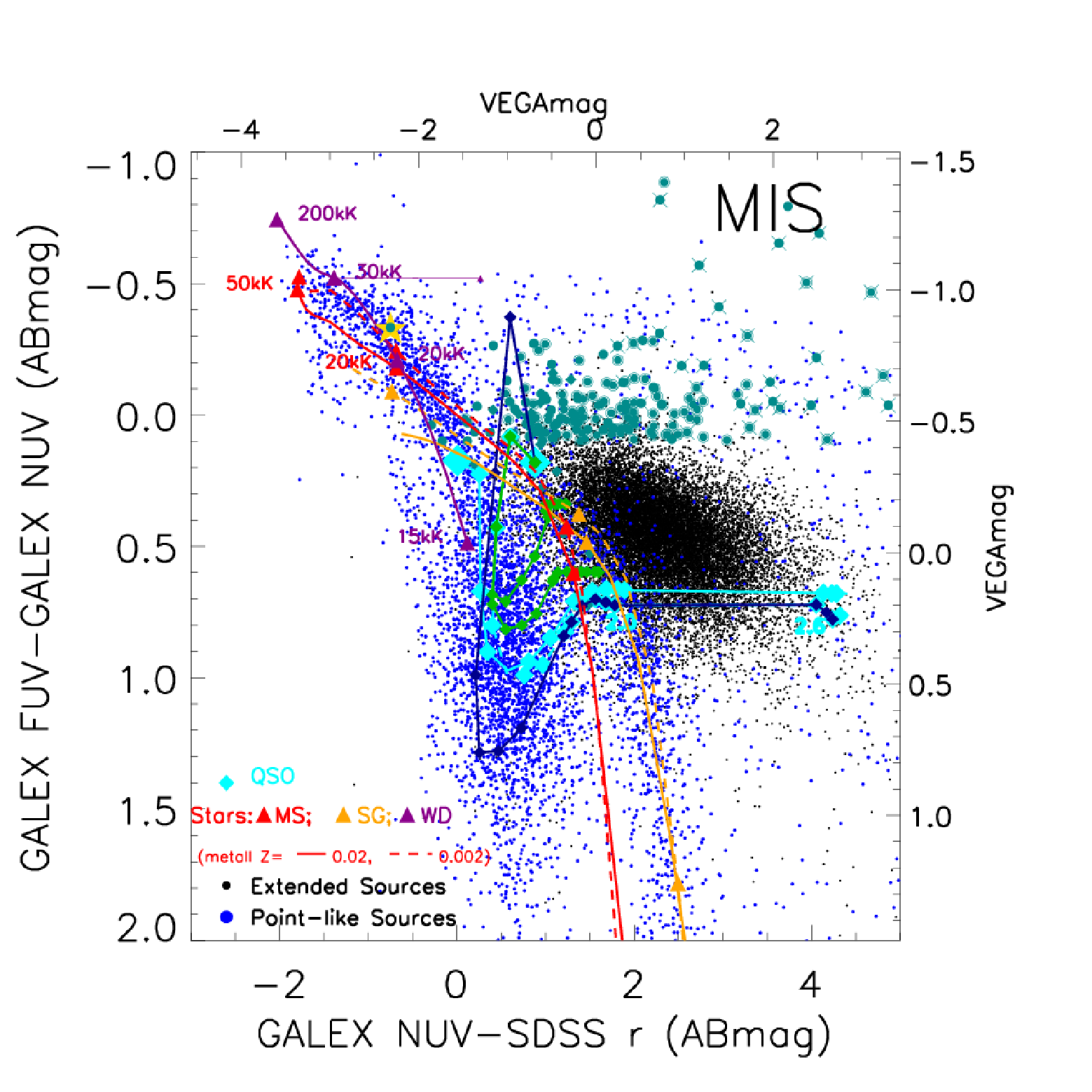}
\caption{\small FULL VERSION AVAILABLE FROM AUTHOR'S WEB SITE: http:\/\/dolomiti.pha.jhu.edu\/papers\/2009\_AJ\_Extreme\_UV\_QSOs.pdf 
Color-color diagram showing the catalog of GR3-MIS UV sources  of Bianchi (2008),
with blue dots for pointlike sources and black dots for extended sources (essentially galaxies). 
Our UV-blue QSOs are shown with teal dots (circled for extended sources, and marked with an X when the FUV-NUV error is $>$.15).
 Synthetic colors for QSO templates are shown with diamonds 
(cyan: standard templates, 
dark blue: templates with  3$\times$ enhanced Ly$\alpha$ emission,
dark green:  templates with F$_{\lambda}$$\sim$${\lambda}^{-0.6}$ and F$_{\lambda}$$\sim$${\lambda}^{-1.2}$ in EUV. 
The redshift values marked by the diamonds are z=0 (largest cyan diamonds, near the center),
0.2,0.4,0.6,1.,1.2,1.4,1.6,1.8,2. (labelled),2.2, 2.4,2.6 (labelled),3.0.
Stellar sequences are shown (red, yellow, and purple triangles for log~g=3, 5 and 9), with \Teff values marked.  
A reddening arrow for E(B-V)=0.3 is shown on a WD (\Teff=30kK) model point.  
The yellow star is the stellar object misclassified by the SDSS pipeline as a QSO. 
The majority of UV-normal QSO follows the template tracks, below and to the left of the stellar sequence.
\label{f_ccd} }
\end{figure}

\begin{figure}
\includegraphics[scale=.75]{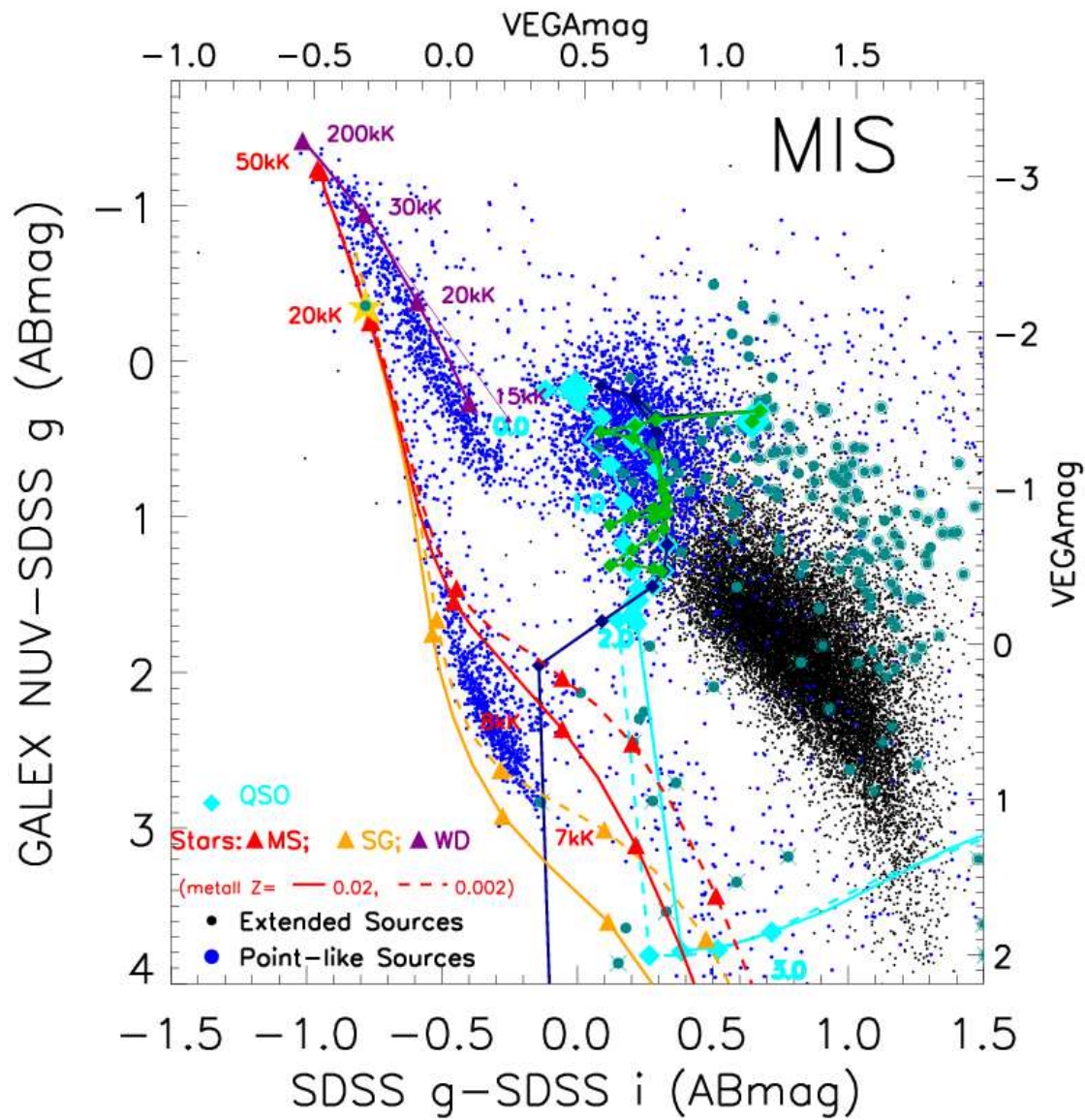}
\caption{\small  FULL VERSION AVAILABLE FROM AUTHOR'S WEB SITE: http:\/\/dolomiti.pha.jhu.edu\/papers\/2009\_AJ\_Extreme\_UV\_QSOs.pdf
Color-color diagram including the g-i color. Symbols as in previous figure.  In this plot, the NUV-g color separates
the high and low redshift QSOs (redshift values marked in cyan along the template, z=0 is the large diamond at the top of the sequence. 
The cluster of pointlike sources close to the low-z QSO template colors are normal QSOs.  
The extended sources among our UV-blue QSOs have g-i redder than QSOs templates, and in the g-i color range of galaxies, 
 reflecting the contribution from the underlying galaxy, 
but they are bluer than the galaxies in NUV-g. This diagram further separates the single  hot stars from the UV-blue QSOs 
(note again the location of the spectroscopically misclassified object, plotted as a yellow star). \label{f_ccd2} }
\end{figure}

\begin{figure} 
\includegraphics[scale=.55]{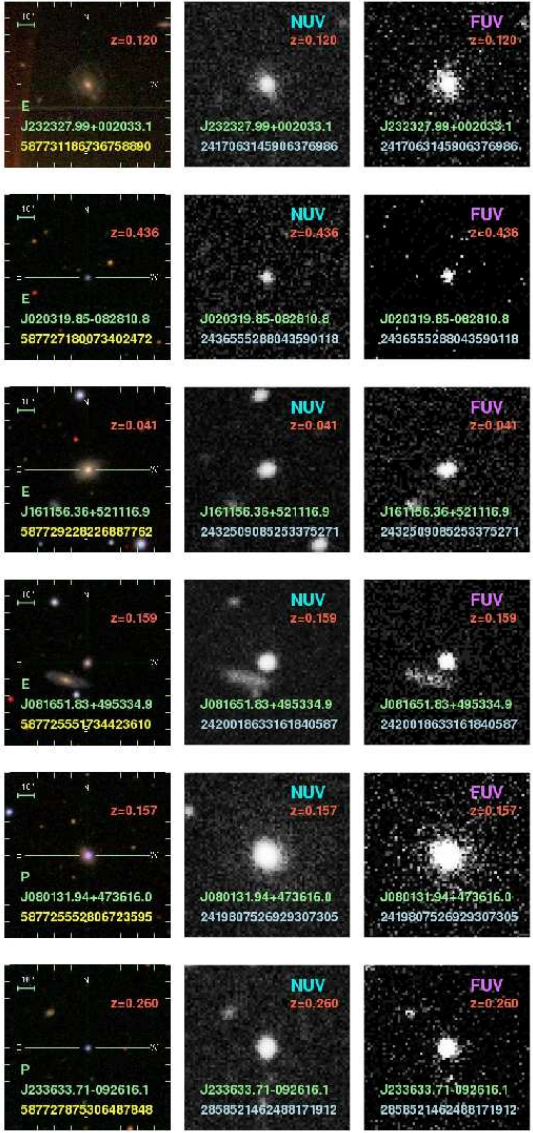}
\caption{\small  FULL VERSION AVAILABLE FROM AUTHOR'S WEB SITE: http:\/\/dolomiti.pha.jhu.edu\/papers\/2009\_AJ\_Extreme\_UV\_QSOs.pdf  ~~~ 
Sample imaging of our UV-blue QSOs. Each row shows one object.  
Columns left to right: First: Color-composite SDSS image (resolution  $\approx$ 1.4 \as),
Second and third: 
GALEX NUV and FUV image respectively (resolution 4.2~\as ~FUV / 5.3~\as ~NUV).
The top 4 sources are classified as extended by the SDSS pipeline, the lower two objects as pointlike.
\label{f_fchart_new} } 
\end{figure}

\begin{figure} 
\includegraphics[scale=.8]{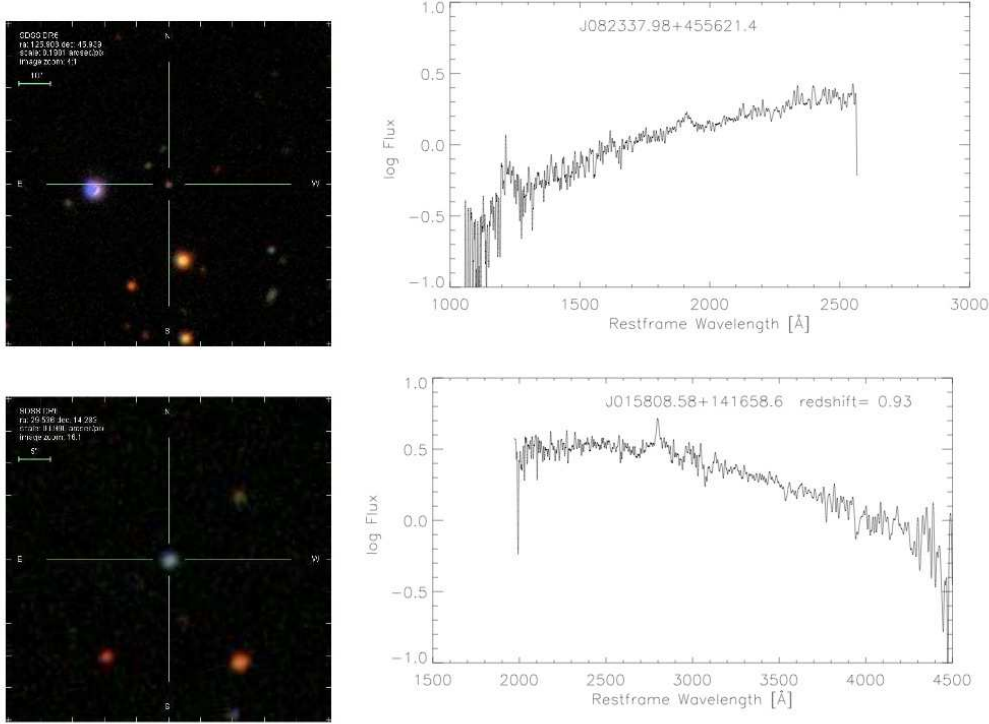}
\caption{Two puzzling objects in the sample. Top: a ``UV-blue'' QSO with a red spectrum. The blue star nearby  
is further away than the match radius used of 4\as. 
Bottom: the only object in the 
sample with redshift near to 1; its classification as a QSO is doubtful. \label{f_oddball} }
\end{figure}

\begin{figure}  
\includegraphics[scale=.7,angle=90.]{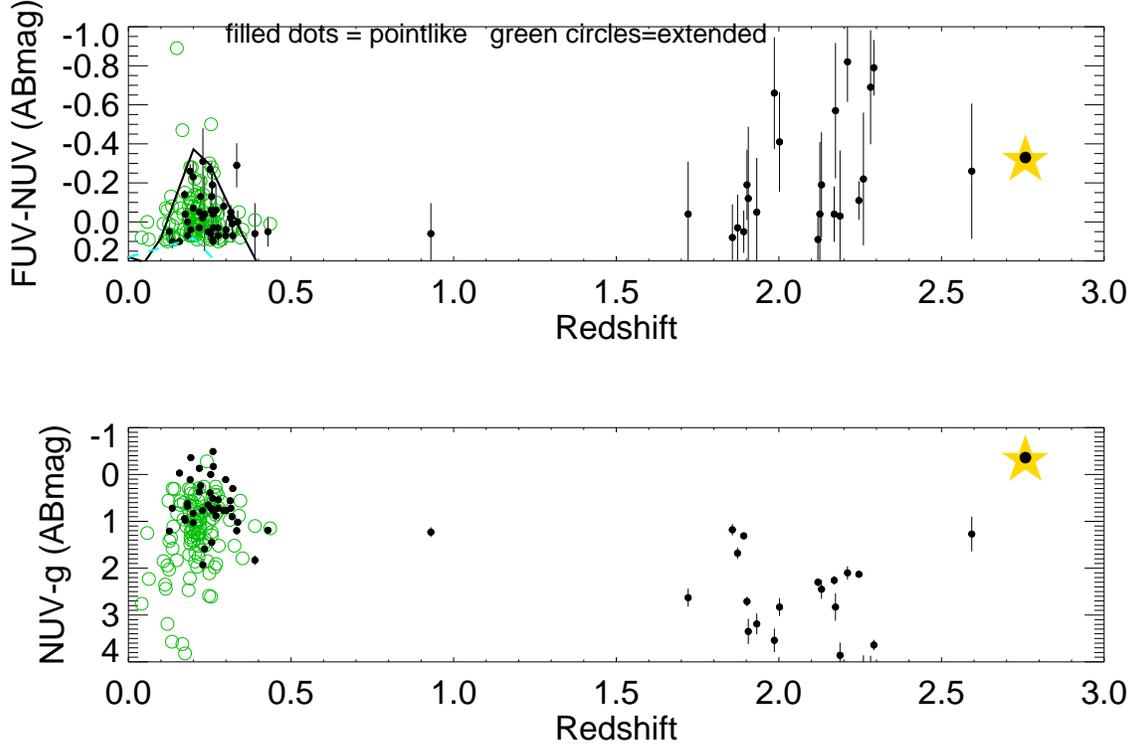}
\caption{\small The measured FUV-NUV color and NUV-g color  versus redshift in the sample 
of UV-blue QSOs. 
Black dots indicate point-like sources (at the SDSS resolution), and green/grey circles extended objects.
The line in the top plot, visible for redshift $\sim$0.1-0.4, 
is the template with 3$\times$~ enhanced Ly$\alpha$. 
The standard templates have redder colors, below the plot range. 
Observed colors are not dereddened. Applying reddening corrections (assuming foreground MW dust with R$_V$=3.1)
makes the FUV-NUV color more negative but insignificantly 
(dots would be higher by an amount about the size of the symbol at most) and decrease the NUV-g color by up to 0.2~mags. 
 The yellow/grey star  marks the source reclassified by us as a hot star, for the object at redshift z=0.93 see text.
The lack of objects in the redshift range 0.5-1.7 is consistent with Ly$\alpha$ being in the NUV band in 
this range (Fig. 1).  The object at redshift z=2.6 has a very red optical spectrum (Fig. \ref{f_oddball}). 
\label{f_uvcolor} }
\end{figure}

\begin{figure} 
\includegraphics[scale=.7]{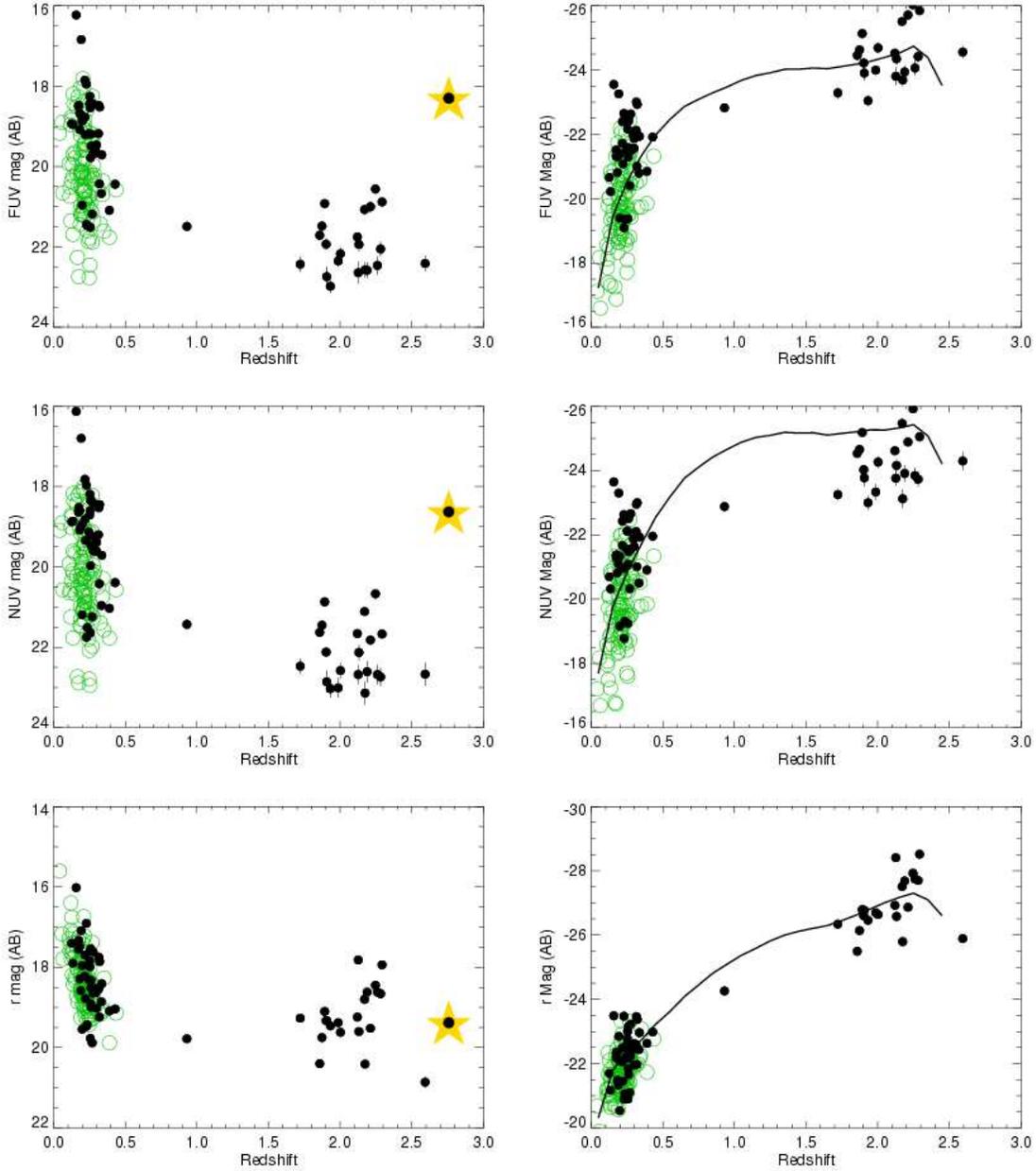}
\caption{\small Left panels: FUV, NUV, and {\it r-}band magnitudes 
(observed, not dereddened) versus redshift. Black dots are 
point-like sources, green/grey circles are extended objects.
Among the low redshift QSOs, pointlike objects tend to be  brighter than extended ones 
in both FUV and NUV, 
while they are similarly spread in {\it r-}band magnitude. 
Right panels: Absolute magnitudes, dereddened for foreground MW extinction. The solid line is the mean values
of the UV-normal QSOs. 
 \label{f_mags}}  
\end{figure}

\begin{figure} 
\includegraphics[scale=.8,angle=90.]{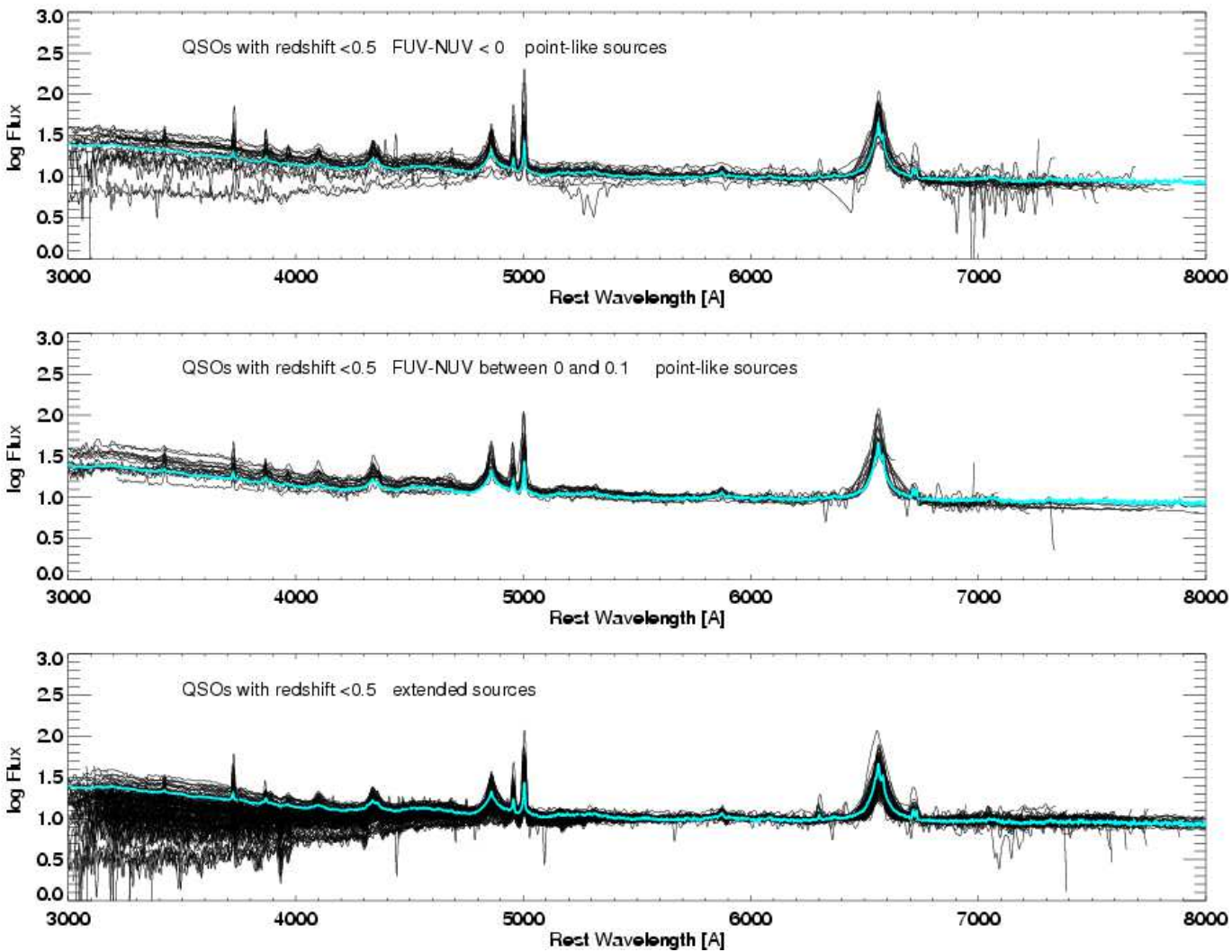}
\caption{The visible spectra of the low redshift QSOs. 
The fluxes (F$_{\lambda}$) have been scaled to a common value 
in the range 6000-6500\AA. Because of the large number of objects,
we plot the pointlike sources in two groups, separated by FUV-NUV color. 
While the general spectral features are well represented by the
template (cyan), stronger emission lines (especially H$\alpha$) and 
spectral slopes bluer than the template are observed in most cases (note the log scale). 
The extended objects are plotted in the bottom panel. Most have a redder slope than the template, and there
is a mix of broad and narrow lines.  
\label{f_sp_stack_loz} } 
\end{figure}

\begin{figure} 
\includegraphics[scale=.89]{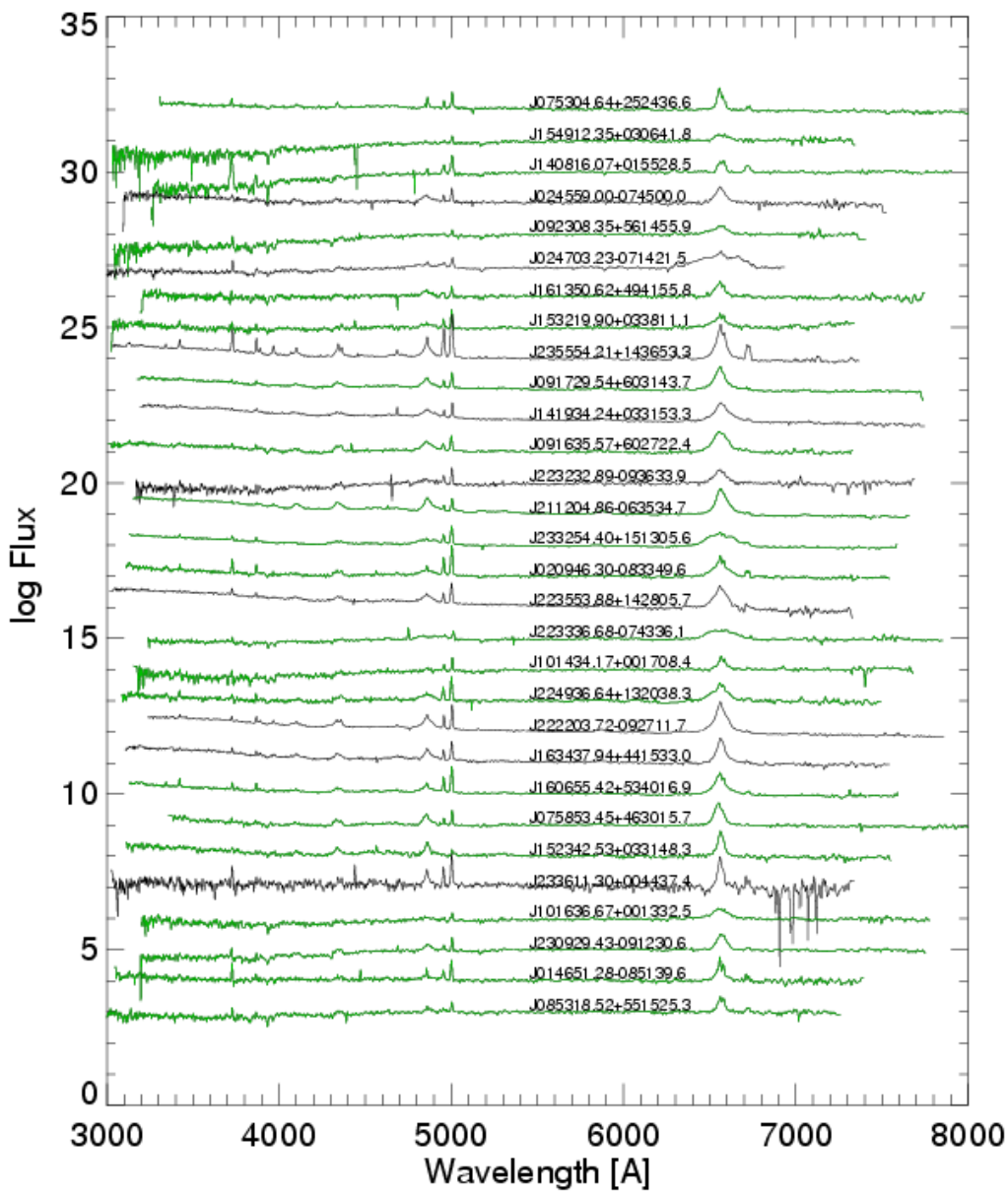}
\caption{\small  Sample spectra of low-redshift UV-blue QSOs, labelled by the GALEX IAU identifier. Fluxes are  scaled to have
a constant offset at 6000-6500\AA. Spectra of extended objects are plotted in green/grey, and of pointlike sources in black.
The order is FUV-NUV bluer to redder, top to bottom, but the spectral features, especially the slope, do
not show any obvious trend with UV color.  \label{f_sp_loz} }
\end{figure}

\begin{figure}  
\includegraphics[scale=.85]{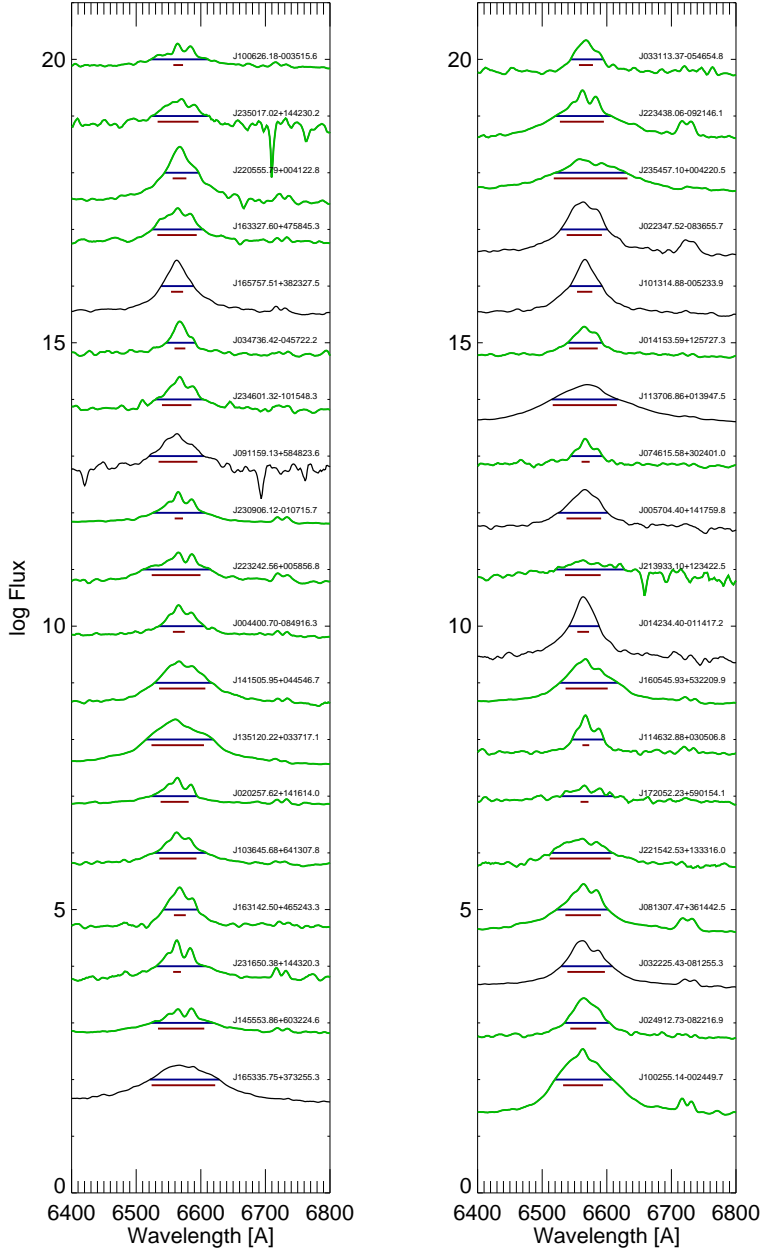}
\caption{\small  The H$\alpha$ line of about one third of the low-z objects (black: pointlike, green: extended sources).
Profiles vary from very broad to narrow, to broad profiles with superimposed narrow components. 
 The line width measured with our code (see text) is shown with a
blue horizontal line, the width  from the SDSS 
 pipeline with a red line (plotted lower for clarity).Fluxes (F$_{\lambda}$) are offset for clarity.
\label{f_hawidth} }
 \end{figure}

\begin{figure}  
\includegraphics[scale=.7]{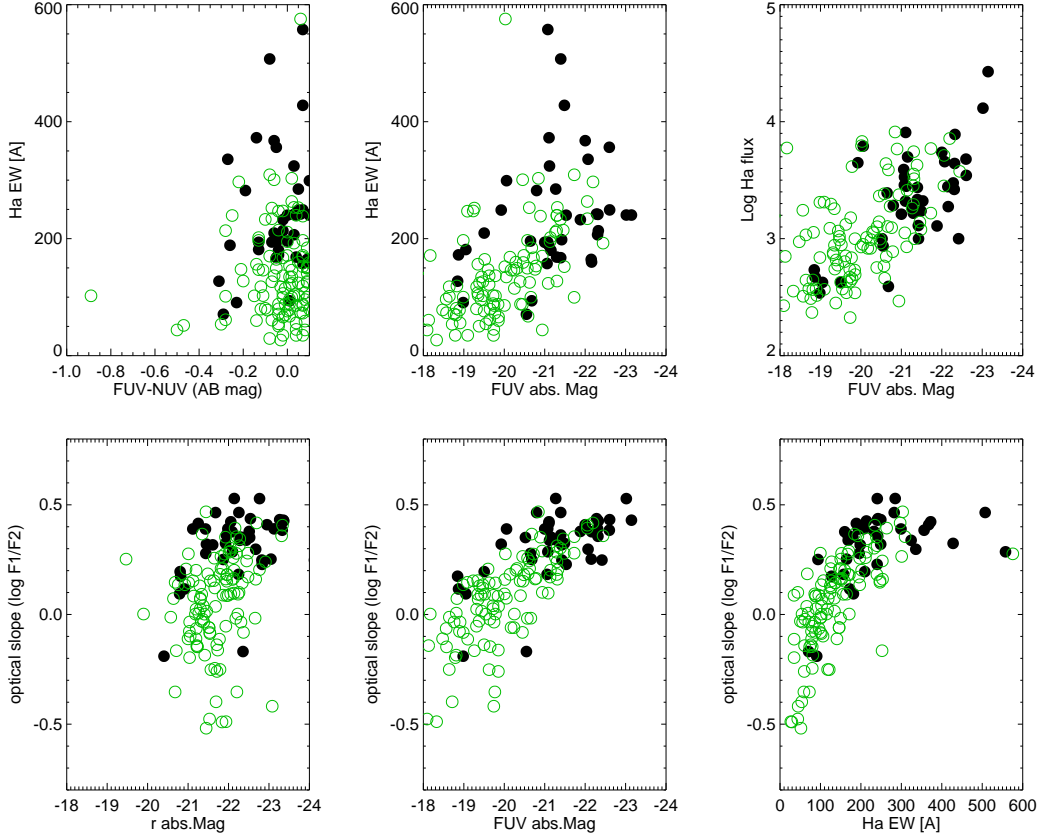} 
\caption{H$\alpha$ emission and optical spectral slope F$_{\lambda}1$(3500-3700\AA)/F$_{\lambda}$2(6000-6400\AA) of the low 
 redshift QSOs show correlation with the UV absolute magnitude, but not with the
 FUV-NUV color or optical absolute magnitude. 
Spectral slope, and to a lesser extent the H$\alpha$ flux and EW, differ
between pointlike (black dots) and extended (green/grey circles) samples. The  H$\alpha$ flux  is in units of 
10$^{-17}$ergs cm$^{-2}$ s$^{-1}$.  
If we restrict the sample to redshift z=0.15-0.25 
(where Ly$\alpha$ is centered in the FUV filter) the correlations in 
the right and middle panels become tighter. \label{f_ha} }
\end{figure}

\begin{figure}  
\includegraphics[scale=1.]{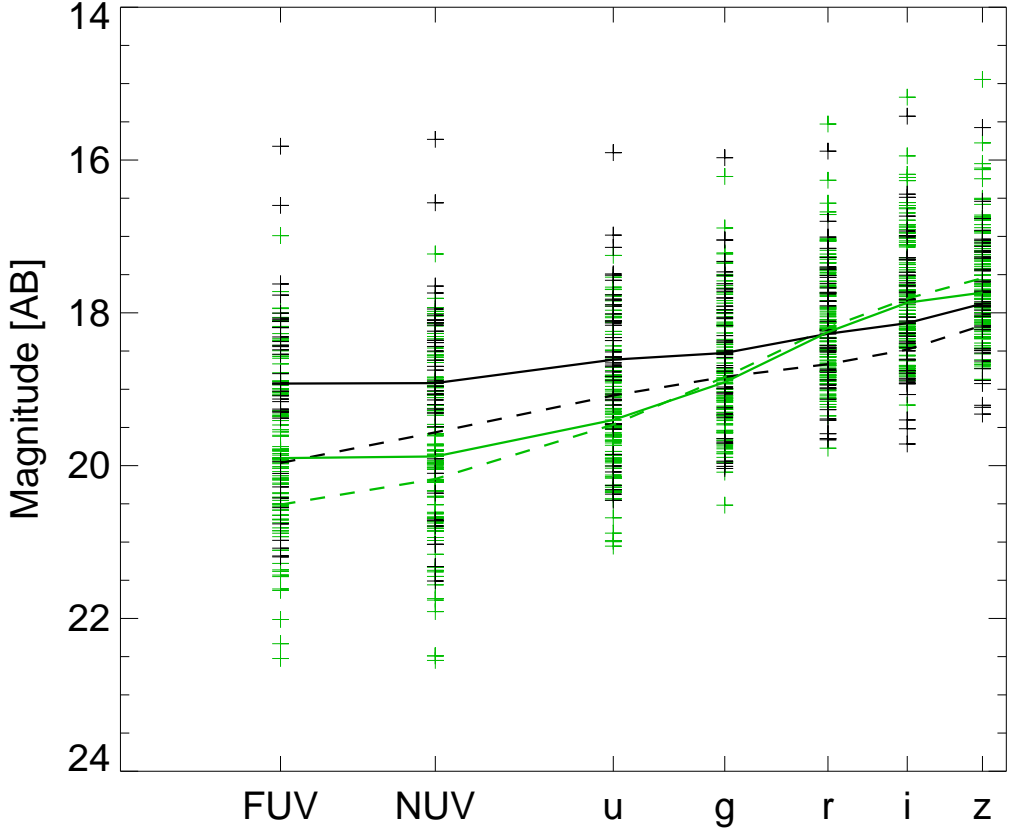}
\caption{\small  GALEX and SDSS magnitudes for the low-redshift sample, 
and median magnitudes as solid lines (black=pointlike, green=extended sources). 
Magnitudes are plotted at the $\lambda$$_{eff}$ of each filter, on a logarithmic wavelength scale.  
Dashed  lines are  the median for the comparison sample of 
UV-normal  QSOs. 
Ly$\alpha$ lies within the FUV channel for these objects and will contribute to the FUV-NUV color. 
 In FUV-NUV the greatest difference is seen, consistent with our sample selection.  
For both UV-blue and UV-normal samples we simply averaged all QSOs within the same redshift range. 
The number of objects across the redshift range however is distributed non uniformly for each sample; if we eliminate the weight
of the relative  number of objects and combine median values in small redshift bins, the curves change very little, and the general trend
is the same. Because average magnitudes vary with redshift, average properties of samples vary according to how the sample
is defined. Therefore small differences should not be overinterpreted but we believe that the general trend is robust.  
\label{f_sedLowZ} }
\end{figure}

\begin{figure}   
\includegraphics[scale=.8]{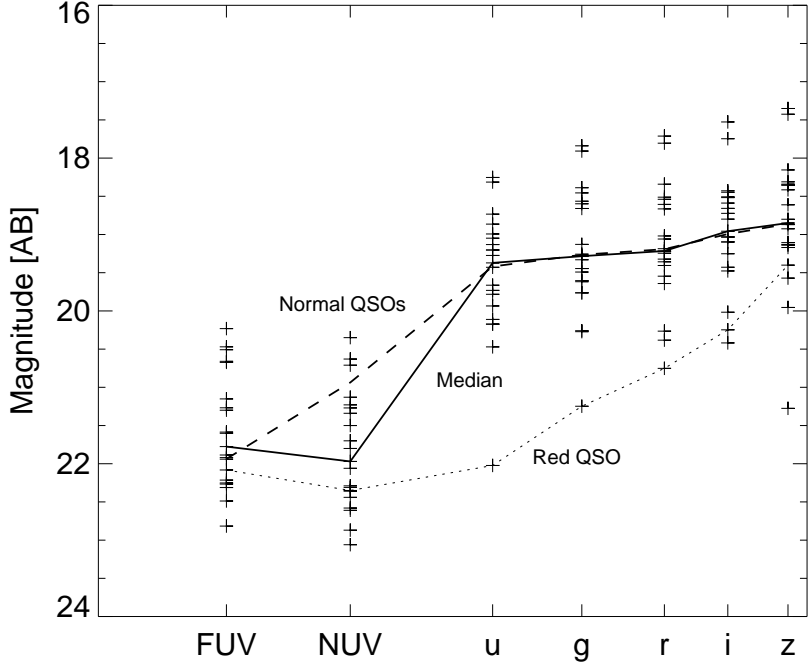}
\caption{\small   GALEX and SDSS average magnitudes for the high redshift sample. The Lyman limit
lies between the NUV and u bands in this redshift range. The line is
the median values for each and the dotted line is the one discrepant QSO 
with a very red optical spectrum. The dashed line is the average from a sample
of UV-normal QSOs within the same redshift range (1.7-2.4). 
Photometry of each object has been corrected for interstellar extinction using 
\ebv~ given in Table 1; the same correction was applied to the UV-normal sample before deriving the average. 
A plot without extinction 
correction applied is qualitatively very similar, shifted slightly towards fainter values,
 especially in UV, but the relative differences remain the same. 
Magnitudes are plotted at the $\lambda$$_{eff}$ of each filter, on a logarithmic wavelength scale.   \label{f_jh1} }
\end{figure}

\begin{figure}   
\includegraphics[scale=.75]{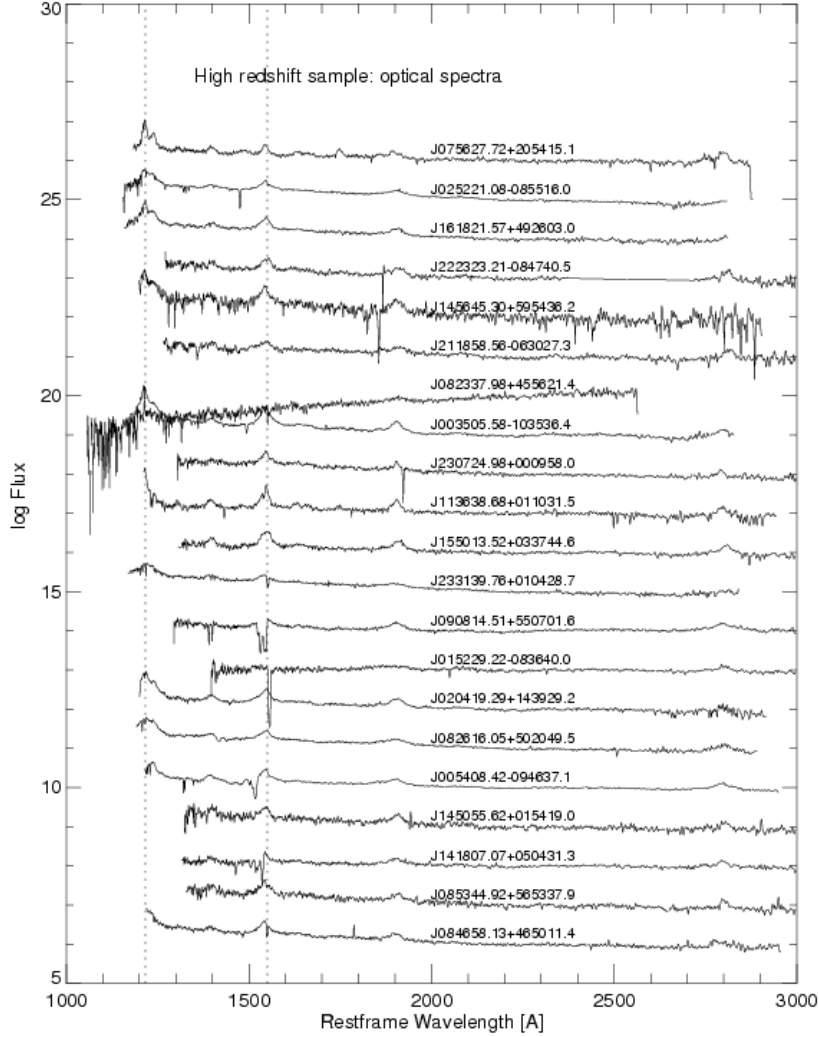}
\caption{\small  The optical spectra of the high-z sample, plotted in the restframe wavelength.
The spectra (in F$_{\lambda}$) are scaled so to have a constant offset at 2000-2500\AA.
They are ordered (top to bottom) by FUV-NUV (bluer to redder), and labelled with their GALEX IAU identifier.
Only for a few objects Ly$\alpha$ is included in the spectrum,
 at the edge of the observed range. 
\label{f_sp_hiz} }
\end{figure}

\begin{figure} %
\includegraphics[scale=.9]{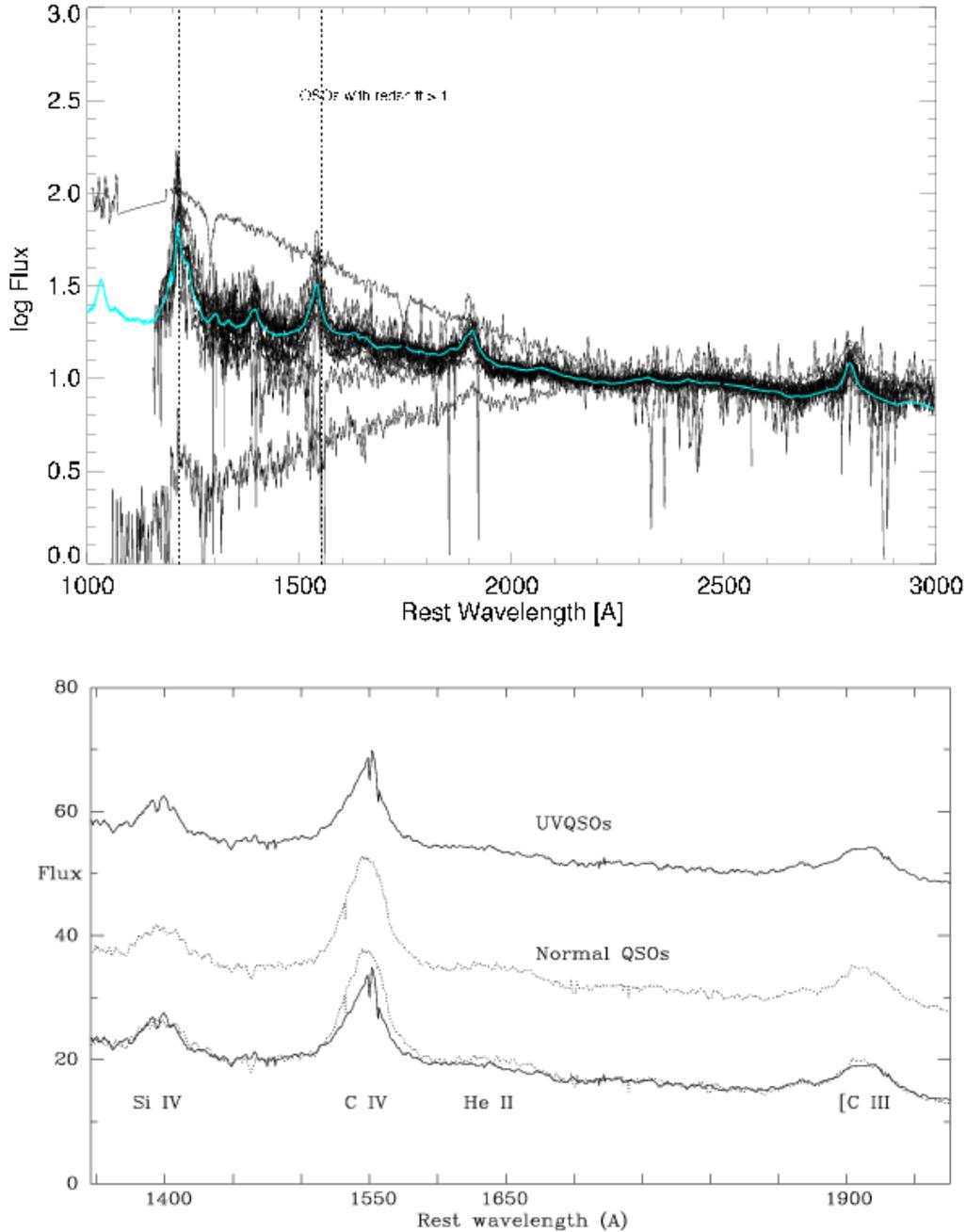}
\caption{\small The visible spectra of the high redshift QSOs. 
Top: fluxes (F$_{\lambda}$)  have been scaled to a common value %
 in the range 2000-2500\AA. 
The ``hottest'' spectrum is the hot star, misclassified by the SDSS
pipeline as a QSO of redshift z=2.7: in the observed wavelength scale, the absorptions are the
Balmer lines. Vertical  dotted lines mark 
Ly$\alpha$ and \ion{C}{4}$\lambda$1550 positions. The cyan spectrum is the standard QSO template. 
 The extremely red spectrum  is discussed separately (Fig \ref{f_oddball}). 
Bottom: Averaged optical spectra of our UV-blue sample, and UV-normal comparison sample 
in the same redshift range. The CIV doublet is different. 
For Ly$\alpha$ no conspicuous difference is seen, but this line is available only for few
``UV-blue'' QSOs, making the comparison not significant. \label{f_sp_stack_hiz}}
 \end{figure}

\begin{figure} 
\includegraphics[scale=.6,angle=90]{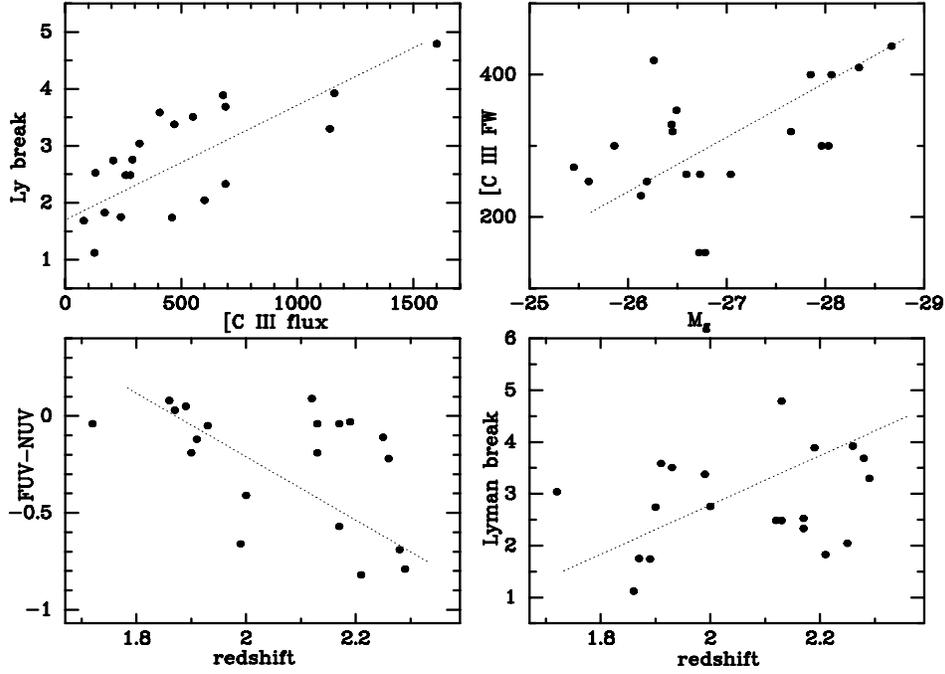}
\caption{\small   Quantities which show trends with UV data in the high redshift sample.
The Lyman break value is the mean of the two GALEX magnitudes minus
the mean of the 5 SDSS magnitudes. The dotted lines are linear fits to 
the points. The red QSO (see Fig  \ref{f_oddball}) is very discrepant in the lower two 
plots and is off-scale and not fitted by the line. FUV-NUV is in AB magnitudes. 
The full width (FW, in \AA) plotted in the upper-right panel is measured at 10\% of the peak flux
above the local continuum, by line profile fitting. \label{f_jh2} }
\end{figure}

\begin{figure}  
\includegraphics[scale=.55,angle=90.]{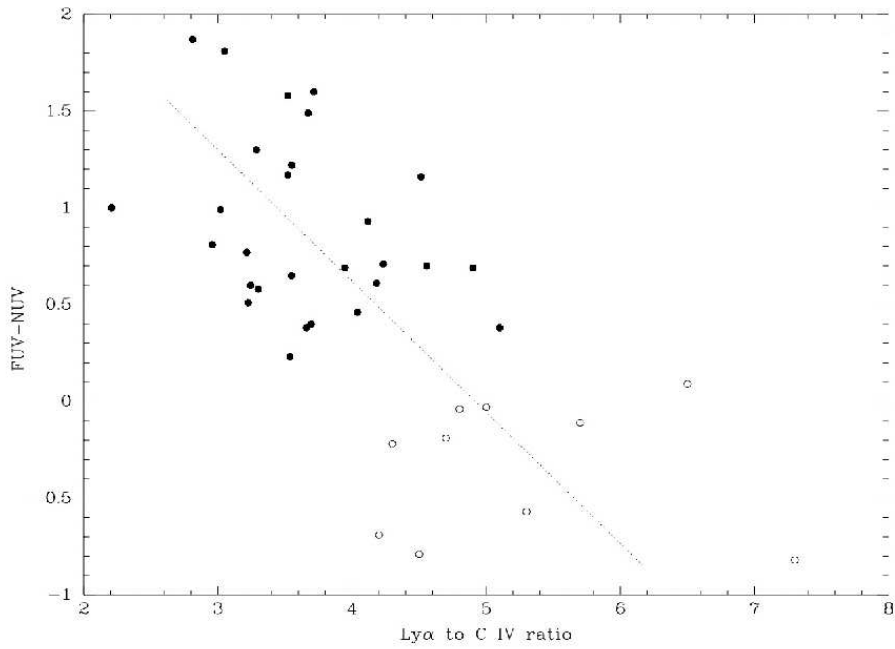}
\caption{\small  Ratio of Ly$\alpha$+NV to CIV emission for the UV sample (circles) and a comparison sample
with similar redshift but FUV-NUV$>$0.1 (filled dots). Although the sample is very limited, 
the UV-blue QSOs tend to have a higher ratio, suggestive that collisions may be  more relevant.
The line is a linear fit.
 \label{f_ratio}}
\end{figure}

For both the low-redshift QSOs where Ly$\alpha$ may be stronger (possibly up to 3$\times$) than in
typical QSO templates, and the high-redshift QSOs where deep Lyman valley absorption  may occur,
UV spectroscopy is needed for a conclusive explanation of the FUV-NUV color, and to assess
whether these are similar to some known objects.
 Our analysis showed that the GALEX photometry provides a unique sieve to select these
UV-blue QSOs, whose optical properties are not unusual. 
 The analysis of this limited spectroscopic
sample, and its average photometric properties, also provides useful information to separate these QSOs
 from stellar binaries with a  hot WD in our larger samples of photometric candidates 
(e.g. Bianchi et al. 2007, 2008), as shown in Figs \ref{f_ccd} and  \ref{f_ccd2}. 
The contamination of these objects in stellar samples may
be very significant at faint magnitudes, because the density of Milky Way hot WDs, extracted from GALEX catalogs,
at MIS depth (UV mag $\sim$ 22.7 ABmag) is much lower than that of QSOs (Bianchi e al. 2007, 2008).

\acknowledgments

We are very grateful to Vahram Chavushyan,  Lucio Buson  and Sebastien Heinis for discussions
at different stages of this work, and to the anonymous referee for many comments which
led to useful clarifications and improved the paper. 
More information and related papers are available at the author's web site at
\url{http://dolomiti.pha.jhu.edu}. 
GALEX (Galaxy Evolution Explorer) is a NASA Small Explorer, launched in April 2003.
We gratefully acknowledge NASA's support for construction, operation,
and science analysis of the GALEX mission,
developed in cooperation with the Centre National d'Etudes Spatiales
of France and the Korean Ministry of 
Science and Technology. 
The data presented in this paper were obtained from the Multimission Archive at the Space Telescope Science Institute (MAST). 
STScI is operated by the Association of Universities for Research in Astronomy, Inc., under NASA contract NAS5-26555. Support for
 MAST for non-HST data is provided by the NASA Office of Space Science via grant NAG5-7584 and by other grants and contracts.



{\it Facilities:}  \facility{GALEX},
\facility{Sloan}, \facility{HST (STIS)}

\newpage

\clearpage

\begin{deluxetable}{lrrrrrrrrrrrrrrrrlll}
\tabletypesize{\tiny}
\rotate
\tablecaption{The UV-blue QSO sample (full table in electronic version) \label{t_sample} }
\tablewidth{0pt}
\tablehead{\colhead{GALEX IAU ID}&\colhead{RA (deg)}&\colhead{Dec(deg)}&\colhead{FUV (AB)}&\colhead{NUV (AB)}&\colhead{u (AB)}&\colhead{g (AB) }&\colhead{r  (AB)}&\colhead{i (AB)}&\colhead{z (AB)}&\colhead{redshift}&\colhead{E(B-V)}&\colhead{Comment}}
\startdata
\cutinhead{High Redshift QSOs (z$>$1) }\\
GALEX J075627.72+205415.1 &  119.1154984 &   20.9041836 &  21.00$\pm$  0.15 &  21.82$\pm$  0.14 &  20.23$\pm$  0.09 &  19.72$\pm$  0.02 &  19.52$\pm$  0.03 &  19.22$\pm$  0.04 &  19.22$\pm$  0.15 & 2.21 & 0.06 & pointlike\\
GALEX J025221.08-085516.0 &   43.0878540 &   -8.9211193 &  20.88$\pm$  0.09 &  21.67$\pm$  0.11 &  18.56$\pm$  0.03 &  18.03$\pm$  0.01 &  17.94$\pm$  0.01 &  17.85$\pm$  0.02 &  17.50$\pm$  0.04 & 2.29 & 0.05 & pointlike\\
GALEX J161821.57+492603.0 &  244.5898736 &   49.4341532 &  22.05$\pm$  0.18 &  22.74$\pm$  0.23 &  19.23$\pm$  0.04 &  18.64$\pm$  0.01 &  18.66$\pm$  0.01 &  18.63$\pm$  0.02 &  18.38$\pm$  0.05 & 2.28 & 0.02 & pointlike\\
GALEX J222323.21-084740.5 &  335.8467089 &   -8.7945764 &  22.35$\pm$  0.14 &  23.01$\pm$  0.25 &  19.52$\pm$  0.05 &  19.47$\pm$  0.02 &  19.38$\pm$  0.02 &  19.14$\pm$  0.03 &  19.00$\pm$  0.10 & 1.99 & 0.05 & pointlike\\
GALEX J145645.30+595436.2 &  224.1887297 &   59.9100553 &  22.57$\pm$  0.19 &  23.14$\pm$  0.29 &  20.52$\pm$  0.08 &  20.31$\pm$  0.03 &  20.41$\pm$  0.04 &  20.44$\pm$  0.07 &  19.97$\pm$  0.15 & 2.17 & 0.01 & pointlike\\
GALEX J211858.56-063027.3 &  319.7439988 &   -6.5075842 &  22.17$\pm$  0.17 &  22.58$\pm$  0.19 &  19.97$\pm$  0.06 &  19.75$\pm$  0.02 &  19.62$\pm$  0.02 &  19.48$\pm$  0.04 &  19.27$\pm$  0.11 & 2.00 & 0.11 & pointlike\\
GALEX J082337.98+455621.4 &  125.9082468 &   45.9392840 &  22.41$\pm$  0.19 &  22.67$\pm$  0.29 &  22.22$\pm$  0.80 &  21.40$\pm$  0.23 &  20.86$\pm$  0.14 &  20.33$\pm$  0.09 &  19.46$\pm$  0.11 & 2.59 & 0.04 & pointlike\\
GALEX J003505.58-103536.4 &    8.7732425 &  -10.5934493 &  22.46$\pm$  0.23 &  22.68$\pm$  0.25 &  19.35$\pm$  0.04 &  18.57$\pm$  0.01 &  18.62$\pm$  0.01 &  18.49$\pm$  0.01 &  18.20$\pm$  0.05 & 2.26 & 0.03 & pointlike\\
GALEX J230724.98+000958.0 &  346.8540794 &    0.1661096 &  21.93$\pm$  0.15 &  22.12$\pm$  0.10 &  19.40$\pm$  0.06 &  19.41$\pm$  0.02 &  19.33$\pm$  0.03 &  19.04$\pm$  0.03 &  19.23$\pm$  0.17 & 1.90 & 0.04 & pointlike\\
GALEX J113638.68+011031.5 &  174.1611550 &    1.1754105 &  21.94$\pm$  0.18 &  22.13$\pm$  0.20 &  19.83$\pm$  0.05 &  19.68$\pm$  0.02 &  19.60$\pm$  0.03 &  19.47$\pm$  0.04 &  19.17$\pm$  0.11 & 2.13 & 0.02 & pointlike\\
GALEX J155013.52+033744.6 &  237.5563540 &    3.6290531 &  22.74$\pm$  0.25 &  22.86$\pm$  0.27 &  19.54$\pm$  0.05 &  19.51$\pm$  0.02 &  19.33$\pm$  0.02 &  18.93$\pm$  0.03 &  18.76$\pm$  0.09 & 1.91 & 0.10 & pointlike\\
GALEX J233139.76+010428.7 &  352.9156781 &    1.0746313 &  20.56$\pm$  0.07 &  20.67$\pm$  0.07 &  18.93$\pm$  0.03 &  18.54$\pm$  0.01 &  18.45$\pm$  0.01 &  18.53$\pm$  0.01 &  18.40$\pm$  0.05 & 2.25 & 0.04 & pointlike\\
GALEX J090814.51+550701.6 &  137.0604696 &   55.1171069 &  22.98$\pm$  0.17 &  23.03$\pm$  0.22 &  20.21$\pm$  0.06 &  19.84$\pm$  0.02 &  19.46$\pm$  0.02 &  19.07$\pm$  0.02 &  18.90$\pm$  0.08 & 1.93 & 0.02 & pointlike\\
GALEX J015229.22-083640.0 &   28.1217418 &   -8.6111176 &  22.43$\pm$  0.19 &  22.47$\pm$  0.19 &  20.27$\pm$  0.10 &  19.84$\pm$  0.03 &  19.27$\pm$  0.03 &  18.84$\pm$  0.03 &  18.83$\pm$  0.11 & 1.72 & 0.02 & pointlike\\
GALEX J020419.29+143929.2 &   31.0803956 &   14.6581219 &  21.07$\pm$  0.11 &  21.11$\pm$  0.09 &  19.11$\pm$  0.03 &  18.85$\pm$  0.01 &  18.80$\pm$  0.01 &  18.61$\pm$  0.02 &  18.43$\pm$  0.06 & 2.17 & 0.05 & pointlike\\
GALEX J082616.05+502049.5 &  126.5668879 &   50.3470827 &  22.58$\pm$  0.20 &  22.61$\pm$  0.27 &  19.19$\pm$  0.04 &  18.75$\pm$  0.01 &  18.62$\pm$  0.01 &  18.60$\pm$  0.02 &  18.37$\pm$  0.05 & 2.19 & 0.04 & pointlike\\
GALEX J005408.42-094637.1 &   13.5350664 &   -9.7769584 &  22.64$\pm$  0.28 &  22.68$\pm$  0.24 &  18.45$\pm$  0.02 &  18.06$\pm$  0.01 &  17.82$\pm$  0.01 &  17.61$\pm$  0.01 &  17.41$\pm$  0.03 & 2.13 & 0.04 & pointlike\\
GALEX J145055.62+015419.0 &  222.7317516 &    1.9052814 &  21.48$\pm$  0.12 &  21.45$\pm$  0.12 &  19.86$\pm$  0.05 &  19.77$\pm$  0.02 &  19.75$\pm$  0.03 &  19.56$\pm$  0.04 &  19.63$\pm$  0.18 & 1.87 & 0.04 & pointlike\\
GALEX J141807.07+050431.3 &  214.5294395 &    5.0753685 &  20.92$\pm$  0.09 &  20.87$\pm$  0.06 &  19.93$\pm$  0.06 &  19.56$\pm$  0.02 &  19.10$\pm$  0.02 &  18.72$\pm$  0.02 &  18.46$\pm$  0.05 & 1.89 & 0.03 & pointlike\\
GALEX J085344.92+565337.9 &  133.4371727 &   56.8938634 &  21.71$\pm$  0.13 &  21.63$\pm$  0.11 &  20.42$\pm$  0.14 &  20.45$\pm$  0.05 &  20.40$\pm$  0.08 &  20.12$\pm$  0.10 &  21.35$\pm$  1.50 & 1.86 & 0.05 & pointlike\\
GALEX J084658.13+465011.4 &  131.7422222 &   46.8364994 &  21.75$\pm$  0.08 &  21.66$\pm$  0.08 &  19.47$\pm$  0.04 &  19.36$\pm$  0.02 &  19.24$\pm$  0.02 &  19.14$\pm$  0.03 &  18.88$\pm$  0.07 & 2.12 & 0.02 & pointlike\\
\cutinhead{Low Redshift QSOs  (z$<$ 1)}\\
GALEX J075304.64+252436.6 &  118.2693129 &   25.4101641 &  19.78$\pm$  0.08 &  20.67$\pm$  0.06 &  19.36$\pm$  0.07 &  18.84$\pm$  0.02 &  18.33$\pm$  0.02 &  17.94$\pm$  0.02 &  17.72$\pm$  0.06 & 0.15 & 0.09 & extended\\
GALEX J154912.35+030641.8 &  237.3014425 &    3.1116111 &  22.45$\pm$  0.28 &  22.95$\pm$  0.28 &  21.62$\pm$  0.57 &  20.34$\pm$  0.06 &  19.01$\pm$  0.03 &  18.48$\pm$  0.03 &  17.96$\pm$  0.09 & 0.25 & 0.13 & extended\\
GALEX J140816.07+015528.5 &  212.0669475 &    1.9245728 &  22.26$\pm$  0.18 &  22.73$\pm$  0.29 &  20.50$\pm$  0.28 &  19.11$\pm$  0.03 &  18.06$\pm$  0.02 &  17.61$\pm$  0.02 &  17.23$\pm$  0.05 & 0.17 & 0.03 & extended\\
\enddata
\end{deluxetable}


\begin{deluxetable}{lcccclccccccl}
\tabletypesize{\tiny}
\rotate
\tablecaption{Objects with Repeated GALEX Observations discrepant by $>$2$\sigma$error \label{t_dup} }
\tablewidth{0pt}
\tablehead{\colhead{GALEX IAU ID}& \colhead{FUV(AB)} & \colhead{NUV(AB)}& \colhead{Date}& \colhead{exp.time (s)} & \colhead{matched GALEX ID} &\colhead{Dist('')}&\colhead{FUV(AB)} &\colhead{NUV(AB)}&\colhead{Date}&\colhead{exp.time(s)}&\colhead{Survey}&\colhead{Comments}}
\startdata
\cutinhead{High Redshift QSOs (z$>$1)}
 J230724.98+000958.0 & 21.93$\pm$0.15  & 22.12$\pm$0.10  & 9/9/2004    & 1697 /3055  &  J230724.98+000958.4 & 0.44 & 21.61$\pm$0.10  & 21.45$\pm$0.08  & 8/24/2003   & 3181 /3181  & MIS &!     \\
 J005408.42-094637.1 & 22.64$\pm$0.28  & 22.68$\pm$0.24  & 9/23/2003   & 1666 /1666  &  J005408.44-094637.7 & 0.71 & 23.14$\pm$0.38  & 21.90$\pm$0.14  & 9/25/2004   & 1648 /1648  & GII &      \\
 J085344.92+565337.9 & 21.71$\pm$0.13  & 21.63$\pm$0.11  & 1/17/2004   & 1694 /1694  &  J085344.84+565338.6 & 1.00 & 22.05$\pm$0.10  & 21.60$\pm$0.09  & 1/17/2004   & 2959 /2959  & MIS &   e E  \\
\cutinhead{Low Redshift QSOs (z$<$1)}
 J075304.64+252436.6 & 19.79$\pm$0.08  & 20.67$\pm$0.06  & 2/14/2006   & 1703 /1703  
 & J075304.67+252436.6 & 0.53 & \nodata & 20.45$\pm$0.06  & 2/15/2006   & 1698 /1698  & MIS &   epb E \\
 J153219.90+033811.1 & 20.92$\pm$0.13  & 21.20$\pm$0.12  & 6/7/2003    & 828  /828   &  J153219.90+033812.3 & 1.21 & 21.45$\pm$0.19  & 20.91$\pm$0.13  & 6/7/2003    & 476  /476   & MIS &     \\
 J223553.88+142805.7 & 19.78$\pm$0.04  & 19.97$\pm$0.03  & 8/13/2005   & 1608 /3007  &  J223553.87+142806.0 & 0.35 & 19.76$\pm$0.01  & 19.64$\pm$0.01  & 8/23/2003   & 31533/31533 & DIS &!   EB \\
 J160655.42+534016.9 & 18.98$\pm$0.02  & 19.11$\pm$0.02  & 6/25/2004   & 2862 /2862  &  J160655.40+534016.7 & 0.29 & 19.71$\pm$0.22  & 19.42$\pm$0.01  & 5/2/2005    & 42   /13341 & DIS &!  b B \\
 J085318.52+551525.3 & 21.88$\pm$0.14  & 21.98$\pm$0.14  & 1/18/2004   & 1694 /1694  &  J085318.57+551525.3 & 0.48 & 22.43$\pm$0.19  & 22.06$\pm$0.14  & 1/3/2006    & 1700 /1700  & GII &     \\
 J014248.83+142126.9 & 21.69$\pm$0.14  & 21.78$\pm$0.16  & 10/4/2004   & 1630 /1630  &  J014248.85+142126.1 & 0.86 & 22.24$\pm$0.16  & 21.94$\pm$0.11  & 11/10/2004  & 1702 /3391  & GII &     \\
 J040446.72-045429.7 & 21.18$\pm$0.10  & 21.24$\pm$0.09  & 11/3/2004   & 2462 /2462  &  J040446.73-045431.3 & 1.60 & 21.17$\pm$0.12  & 20.58$\pm$0.11  & 11/3/2004   & 1535 /1535  & MIS &!  eb EPB \\
  J013928.93-103425.9 & 20.81$\pm$0.08  & 20.85$\pm$0.06  & 10/15/2003  & 1579 /1579  & J013928.89-103426.5 & 0.77 & 20.59$\pm$0.05  & 20.58$\pm$0.04  & 12/13/2004  & 3284 /3753  & GII &!    \\
 J234019.81+005907.9 & 19.18$\pm$0.02  & 19.20$\pm$0.02  & 8/23/2003   & 3145 /3145 & J234019.83+005910.3 & 2.46 & 19.45$\pm$0.04  & 19.19$\pm$0.03  & 9/22/2006   & 1669 /1669  & MIS &!   EP \\
  J235457.10+004220.5 & 18.25$\pm$0.03  & 18.22$\pm$0.02  & 10/22/2006  & 1108 /1108 & J235457.15+004220.3 & 0.72 & 18.34$\pm$0.03  & 18.20$\pm$0.02  & 10/21/2006  & 924  /924   & MIS &    EB \\
 J160545.93+532209.9 & 18.26$\pm$0.02  & 18.22$\pm$0.01  & 7/31/2004   & 1637 /1637  &  J160545.98+532210.8 & 1.00 & 18.08$\pm$0.12  & 17.86$\pm$0.00  & 5/2/2005    & 42   /13341 & DIS &!   PB \\
 J032225.43-081255.3 & 18.93$\pm$0.03  & 18.88$\pm$0.02  & 11/29/2003  & 1698 /1698  &  J032225.37-081255.4 & 0.90 & 18.92$\pm$0.04  & 18.72$\pm$0.02  & 11/30/2003  & 1698 /1698  & MIS &!  e E \\
 J160815.26+524450.8 & 19.46$\pm$0.04  & 19.39$\pm$0.03  & 7/31/2004   & 1635 /1635  &  J160815.21+524451.3 & 0.71 & 19.86$\pm$0.28  & 19.73$\pm$0.01  & 5/3/2005    & 40   /14866 & DIS &!  e EPB \\
 J161156.36+521116.9 & 19.17$\pm$0.04  & 19.09$\pm$0.02  & 7/31/2004   & 1635 /1635  &  J161156.35+521117.2 & 0.36 & 19.10$\pm$0.04  & 19.00$\pm$0.02  & 7/31/2004   & 1631 /1631  & NGS &!  eb B \\
 J005057.44+143753.7 & 21.34$\pm$0.11  & 21.26$\pm$0.10  & 9/25/2003   & 1687 /1687  &  J005057.45+143753.8 & 0.21 & 21.70$\pm$0.11  & 21.40$\pm$0.09  & 8/30/2003   & 1967 /1967  & MIS &     \\
 J005328.80-085754.8 & 18.44$\pm$0.02  & 18.36$\pm$0.01  & 9/16/2003   & 1602 /1602  &  J005328.79-085754.1 & 0.70 & 18.33$\pm$0.02  & 18.22$\pm$0.01  & 9/23/2003   & 1666 /1666  & MIS &!    \\
 J163625.47+421346.1 & 18.44$\pm$0.02  & 18.35$\pm$0.02  & 5/23/2004   & 1702 /1702  &  J163625.48+421346.8 & 0.72 & 18.31$\pm$0.03  & 18.26$\pm$0.01  & 7/8/2004    & 1195 /1586  & DIS &!    \\
 J233633.71-092616.1 & 18.44$\pm$0.02  & 18.34$\pm$0.01  & 9/2/2006    & 2189 /2189 &  J233633.72-092616.9 & 0.82 & 18.42$\pm$0.02  & 18.24$\pm$0.01  & 9/27/2005   & 1973 /2859  & MIS &!  eb EB \\
\enddata
\tablecomments{Explanation of comments: "!" denotes magnitude discrepancy (NUV or FUV) greater than 3$\sigma$,  "E" denotes that the EDGE artifact flag is set,  "P" denotes SExtractor flag 1 set (object has neighbours) 
 "B" denotes SExtractor flag 2 set (object was originally blended with another one). 
Lower case flag codes for original sources, upper case for matched measurements.}
\end{deluxetable}



\begin{deluxetable}{lrrr}
\tabletypesize{\tiny}
\tablecaption{H$\alpha$ measurements (restframe) of the low-redshift QSOs  (full table given in electronic version) \label{t_meas}}
\tablehead{\colhead{GALEX IAU ID}&\colhead{H$\alpha$ Width}&\colhead{EW}&\colhead{log flux} \\
\colhead{ }&\colhead{[\AA]}&\colhead{[\AA]}&\colhead{10$^{-17}$ ergs cm$^{-2}$ s$^{-1}$\AA$^{-1}$} }
\startdata
GALEX J075304.64+252436.6 &    47.3 &   102.3 &  -14.01\\
GALEX J154912.35+030641.8 &   109.0 &    43.9 &  -14.57\\
GALEX J140816.07+015528.5 &    57.2 &    51.7 &  -14.31\\
GALEX J024559.00-074500.0 &    86.1 &   127.3 &  -14.27\\
GALEX J092308.35+561455.9 &   114.1 &    53.7 &  -14.43\\
GALEX J024703.23-071421.5 &    73.3 &    70.8 &  -14.06\\
GALEX J161350.62+494155.8 &    79.0 &   101.5 &  -14.35\\
GALEX J153219.90+033811.1 &    94.1 &    61.4 &  -14.46\\
GALEX J235554.21+143653.3 &    53.4 &   335.7 &  -13.34\\
GALEX J091729.54+603143.7 &    79.0 &   213.7 &  -13.55\\
GALEX J141934.24+033153.3 &    86.0 &   188.7 &  -13.68\\
GALEX J091635.57+602722.4 &    98.8 &   239.6 &  -13.71\\
GALEX J223232.89-093633.9 &    95.7 &    90.7 &  -14.46\\
GALEX J211204.86-063534.7 &    80.0 &   297.0 &  -13.14\\
GALEX J233254.40+151305.6 &   145.7 &   127.6 &  -13.45\\
GALEX J020946.30-083349.6 &    90.4 &   147.6 &  -13.91\\
GALEX J223553.88+142805.7 &    96.3 &   282.3 &  -13.72\\
GALEX J223336.68-074336.1 &   143.8 &   102.3 &  -13.96\\
GALEX J101434.17+001708.4 &    76.8 &    59.9 &  -14.46\\
GALEX J224936.64+132038.3 &   107.9 &   156.3 &  -14.06\\
\enddata
\end{deluxetable}

\begin{deluxetable}{lcc}
\tabletypesize{\tiny}
\tablecaption{Average line measurements of the high redshift sample \label{t_ratios}}
\tablehead{\colhead{ }&\colhead{UV QSOs}&\colhead{UV-Normal QSOs} } 
\startdata
   \# objects        &   10    &        28 \\
    Ly$\alpha$/C IV  &  5.23  &       3.68  \\
    Ly$\alpha$ flux  &  440    &      285 (365)\\ 
     C IV flux    &   90    &       75 (96)\\ 
     C III] flux  &   52    &       41  (52)\\ 
     C IV/ C III] &  1.74   &      1.82   
\enddata
\tablecomments{Values in parentheses for UV-normal QSOs are continuum-corrected. 
Note that the average of the ratios and the ratio of the averages
are not the same because the average fluxes weigh the high values more. Theoretical values
for  Ly$\alpha$/C IV ratio are 6.7 for collisional ionization and 1.8 for photoionization.}
\end{deluxetable}

\end{document}